\documentclass[aps,prd,nofootinbib,showpacs,superscriptaddress,preprint]{revtex4}
\pdfoutput=1
\usepackage[utf8]{inputenc}
\usepackage{epsfig}
\usepackage{amsmath}
\usepackage{slashed} 
\usepackage{amsfonts} 
\usepackage{float} 
\usepackage{amssymb}
\usepackage{color}
\usepackage{pbox}
\usepackage{subfigure}
\usepackage{array,multirow,makecell}
\usepackage[colorlinks,citecolor=blue]{hyperref}
\usepackage{tabularx}
\usepackage{titlesec} 
\usepackage{natbib}
\usepackage{lipsum}
\usepackage{slashed}
\usepackage{tikz}
\usetikzlibrary{snakes}

\def\lsim{\raise0.3ex\hbox{$\;<$\kern-0.75em\raise-1.1ex\hbox{$\sim\;$}}}
\def\gsim{\raise0.3ex\hbox{$\;>$\kern-0.75em\raise-1.1ex\hbox{$\sim\;$}}}
\def\be{\begin{equation}}
	\def\ee{\end{equation}}
\def\bea{\begin{eqnarray}} 
	\def\eea{\end{eqnarray}}

\newcommand{\beq}{\begin{equation}}
	\newcommand{\eeq}{\end{equation}}
\newcommand{\beqn}{\begin{eqnarray}}
	\newcommand{\eeqn}{\end{eqnarray}}

\def\lsim{\raise0.3ex\hbox{$\;<$\kern-0.75em\raise-1.1ex\hbox{$\sim\;$}}}
\def\gsim{\raise0.3ex\hbox{$\;>$\kern-0.75em\raise-1.1ex\hbox{$\sim\;$}}}
\def\be{\begin{equation}}
	\def\ee{\end{equation}}
\def\bea{\begin{eqnarray}}
	\def\eea{\end{eqnarray}}

\titleformat{\paragraph}
{\normalfont\normalsize\bfseries}{\theparagraph}{1em}{}
\titlespacing*{\paragraph}
{0pt}{3.25ex plus 1ex minus .2ex}{1.5ex plus .2ex}
\begin{document}
	
	
	\title{TeV Scale Modified Type-II Seesaw and Dark Matter in a Gauged $U(1)_{\rm B-L}$ Symmetric Model}
	
	\author{Purusottam Ghosh}
	\email{purusottamghosh@hri.res.in}
	\affiliation{Regional Centre for Accelerator-based Particle Physics,
		Harish-Chandra Research Institute, HBNI,
		Chhatnag Road, Jhunsi, Allahabad - 211 019, India}
	
	\author{Satyabrata Mahapatra}
	\email{ph18resch11001@iith.ac.in}
	\affiliation{Department of Physics, Indian Institute of Technology Hyderabad, Kandi, Sangareddy 502285, Telangana, India}

	\author{Nimmala Narendra}
	\email{nnarendra@prl.res.in}
	\affiliation{Theoretical Physics Division, Physical Research Laboratory, Ahmedabad 380009, India}
	
	\author{Narendra Sahu}
	\email{nsahu@phy.iith.ac.in}
	\affiliation{Department of Physics, Indian Institute of Technology Hyderabad, Kandi, Sangareddy 502285, Telangana, India}
	
	\begin{abstract}
		In an endeavor to explain the light neutrino masses and dark matter (DM) simultaneously, we study a gauged $U(1)_{\rm B-L}$ extension of 
		the standard model (SM). The neutrino masses are generated through a variant of type-II seesaw mechanism in which one of the scalar 
		triplets has a mass in a scale that is accessible at the present generation colliders. Three SM singlet right chiral fermions $\chi_{iR}$($i=e,\mu,\tau$) 
		with $\rm B-L$ charges -4, -4, +5 are invoked to cancel the $\rm B-L$ gauge  anomalies and the lightest one among these three fermions becomes 
		a viable DM candidate as their stability is guaranteed by a remnant $\mathcal Z_2$ symmetry to which $U(1)_{\rm B-L}$ gauge symmetry gets 
		spontaneously broken. Interestingly in this scenario, the neutrino mass and the co-annihilation of DM are interlinked through the breaking 
		of $U(1)_{\rm B-L}$ symmetry. 
	Apart from giving rise to the observed neutrino 
			mass and dark matter abundance, the model also predicts exciting signals at the colliders. Especially 
			we see a significant enhancement in the production cross-section of the TeV scale doubly charged scalar in presence of the $Z_{\rm BL}$ gauge boson.
		We discuss all the relevant constraints on model parameters from observed 
		DM abundance and null detection of DM at direct and indirect search experiments as well as the constraints on the $\rm B-L$ gauge boson from 
		recent colliders.	
	\end{abstract}
	\maketitle
	\section{Introduction}
	\label{intro}
	Out of all the lacunae afflicting the Standard Model(SM) of particle physics, the identity of DM and the origin of tiny but nonzero neutrino masses 
	are the most irking ones. It is well established by now, thanks to numerous irrefutable observational evidences from astrophysics 
	and cosmology like galaxy rotation curves, gravitational lensing, Cosmic Microwave Background (CMB) acoustic oscillations etc.~\cite{Bertone:2004pz,Zwicky:1933gu,Rubin:1970zza,Clowe:2006eq,Hinshaw:2012aka,Aghanim:2018eyx}, that a mysterious, non-luminous and 
	non-baryonic form of matter exists called as dark matter (DM) which constitutes almost 85\% of the total matter content and 
	around 26.8\% of the total energy density of the present Universe. In terms of density parameter  $h = \text{(Hubble Parameter)}/(100 \;\text{km} ~\text{s}^{-1} \text{Mpc}^{-1})$, the present DM abundance is conventionally reported as \cite{Hinshaw:2012aka,Aghanim:2018eyx} : $\Omega_{\text{DM}} h^2 = 0.120 \pm 0.001$.   
	But still we have no answer
	to the question what DM actually is, as none of the SM particle has the properties that a DM particle is expected to have. Thus over the years, various 
	beyond SM (BSM) scenarios have been considered to explain the puzzle of DM, with additional field content and augmented symmetry. The most popular among these 
	ideas is something known as the weakly interacting massive particle (WIMP) paradigm. In this WIMP scenario, a DM candidate typically having a mass similar to electroweak (EW) scale and interaction rate analogous to EW interactions can give rise to the correct DM relic abundance, an astounding coincidence referred 
	to as the {\it WIMP Miracle}~\cite{Kolb:1990vq,Arcadi:2017kky}. The sizeable interactions of WIMP DM with the SM particles has many phenomenological implications. 
	Along with giving the correct relic abundance of DM through thermal freeze-out, it also leads to other phenomenological implications like optimistic direct 
	and indirect detection prospects of DM which makes it more appealing. Several direct detection experiments like LUX, PandaX-II and XENON1T \cite{Akerib:2016vxi, Tan:2016zwf, Cui:2017nnn, Aprile:2017iyp,Aprile:2018dbl} and indirect detection experiments like space-based telescopes Fermi-LAT and ground-based 
	telescopes MAGIC~\cite{Ackermann:2015zua,Ahnen:2016qkx} have been looking for signals of DM and have put constraints on DM-nucleon scattering cross-sections 
	and DM annihilation cross-section to SM particles respectively. 
	
	Apart from the identity of DM, another appealing motivation for the BSM is the origin of neutrino masses. Despite compelling evidences for existence of light 
	neutrino masses, from various oscillation experiments~\cite{Fukuda:1998mi,Ahmad:2001an,Abe:2011fz,An:2012eh,Ahn:2012nd} and cosmological data~\cite{Tanabashi:2018oca,Vagnozzi:2017ovm,Giusarma:2016phn,Giusarma:2018jei}, the origin of light neutrino masses is still unknown. The oscillation data is only 
	sensitive to the difference in mass-squareds\cite{Tanabashi:2018oca,Esteban:2018azc}, but the absolute mass scale is constrained to $\sum_i |m(\nu_i)| 
	< 0.12$ eV~\cite{Tanabashi:2018oca} from cosmological data. This also implies that we need new physics in BSM to incorporate the light neutrino masses 
	as the Higgs field, which lies at the origin of all massive particles in the SM, can not have any Yukawa coupling with the neutrinos due to the absence of 
	its right-handed counterpart.
	
	Assuming that the neutrinos to be of Majorana type (which violates lepton number by two units), the origin of the tiny but non-zero neutrino mass is 
	usually explained by the see-saw mechanisms (Type-I~\cite{Minkowski:1977sc,GellMann:1980vs,Mohapatra:1979ia,Schechter:1980gr,Davidson:1987mh}, Type-II~\cite{Mohapatra:1980yp,Lazarides:1980nt,Schechter:1981cv,Wetterich:1981bx,Brahmachari:1997cq} and Type-III~\cite{Foot:1988aq}) which are the ultraviolet 
	completed realizations of the dimension five Weinberg operator $\mathcal{O}_5= y_{ij}\frac{\overline{L_i}L^c_jHH}{\Lambda}$, where $L$ and $H$ are 
	the lepton and Higgs doublets of the SM and $\Lambda$ is the scale of new physics~\cite{Weinberg:1979sa,Ma:1998dn}. In the type-I seesaw heavy singlet 
	RHNs are introduced while in type-II and type-III case, a triplet scalar($\Delta$) of hyper-charge 2 and triplet fermions $\Sigma$
	of hyper-charge 0 are introduced respectively such that new Yukawa terms can be incorporated in the theory. Tuning the Yukawa coupling and the 
	cut-off scale ($\Lambda$) and adopting a necessary structure for the mass matrix, the correct masses and mixings of the neutrinos can be obtained. 
	
	In the conventional type-II seesaw, the relevant terms in the Lagrangian violating lepton number by two units are $f_{ij}\Delta L_{i} L_{j} + 
		\mu \Delta^{\dagger}HH$, where $\Delta$ does not acquire an explicit vacuum expectation value(vev). However, after the electro-weak phase transition, a 
		small induced vev of $\Delta$ can be obtained as: $\langle \Delta \rangle = -\frac{\mu  \langle H \rangle^2}{M^2_{\Delta}}$. Thus for 
		$\mu \sim M_{\Delta} \sim 10^{14}$GeV, one can get $M_{\nu}=f \langle \Delta \rangle \simeq f\frac{\langle H \rangle^2}{M_{\Delta}}$ of order  
		$\mathcal{O}$(0.1)eV for $f\sim 1$.
	
	In an alternative fashion, neutrino mass can be generated in a modified type-II seesaw if one introduces two scalar triplets: $\Delta$  and  
	$\xi$ with $M_{\Delta} \sim \mathcal{O}(10^{3})$TeV and $M_{\xi} \sim$ TeV $<<$ $M_{\Delta}$\cite{McDonald:2007ka} \footnote{See also ref. \cite{Gu:2009hu} for 
		a modified double type-II seesaw with TeV scale scalar triplet.}. In this case imposition of additional $\rm B-L$ gauge symmetry~\cite{Majee:2010ar} 
	allows for $\mu \Delta^{\dagger}HH + f\xi LL + y\Phi^2_{\rm  BL}\Delta^{\dagger} \xi$ terms in the Lagrangian (with proper choice of gauge charges 
	for the scalars) where $\Phi_{\rm  BL}$ is the scalar field responsible for $\rm B-L$ symmetry breaking at TeV scale. As is clear from the Lagrangian 
	terms, once the $\Phi_{\rm BL}$ acquires a vev, it creates a small mixing between $\Delta $ and $\xi$ of the order $\theta \sim \frac{\langle \Phi_{ BL}\rangle^{2}}
	{M^2_{\Delta}} \simeq 10^{-6}$. Thus the coupling of $\xi$ with Higgs becomes extremely suppressed but $\xi L L$ coupling can be large. In this scenario, 
	$\Delta$ being super heavy gets decoupled from the low energy effective theory but $\xi$ can have mass from several hundred GeV to a few TeV and having large 
	Di-lepton coupling can be probed at colliders through the same sign Di-lepton signature ~\cite     
	{Huitu:1996su,Chun:2003ej,Perez:2008ha,Padhan:2019jlc,Dev:2018kpa,Barman:2019tuo,Bhattacharya:2018fus}.

	In this paper we implement such a modified type-II seesaw in a gauged $U(1)_{\rm B-L}$ symmetric model and study the consequences for dark matter, neutrino 
	mass and collider signatures. We introduce three right chiral fermions $\chi_{i_R}$ $(i=e,\mu,\tau)$ with $U(1)_{{\rm B-L}}$ charges $-4, -4, +5$ for 
	cancellation of non-trivial gauge and gravitational anomalies.
	For the details of the anomaly cancellation in a $\rm B-L$ model, please see appendix~\ref{anomaly}.   Interestingly, the lightest one among these three exotic fermions becomes a viable candidate of DM, 
	thanks to the remnant $\mathcal Z_2$ symmetry after $U(1)_{\rm B-L}$ breaking, under which $\chi_{i_R}$ $(i=e,\mu,\tau)$ are odd while all other particles are even. \footnote{ Right-handed neutrinos with $\rm B-L$ charge -1 can also serve the purpose of $\rm B-L$ anomaly cancellation and be viable DM candidate provided one introduces an additional ad hoc $\mathcal Z_2$ symmetry to guarantee their stability. For instance see~\cite{Rodrigues:2018jpv,Borah:2018smz}.} Such an alternative integral $\rm B-L$ charge assignment solution for the additional fermions to achieve anomaly cancellation was first proposed in~\cite{Montero:2007cd} and its phenomenology have been studied in different contexts relating to fermionic DM and neutrino mass in~\cite{Ma:2014qra,Sanchez-Vega:2015qva,Singirala:2017cch,Okada:2018tgy,Asai:2020xnz}\footnote{In an earlier preliminary project~\cite{Mahapatra:2020dgk}, we had studied this model with the same motivation. The present work is an extended version with a detailed study of co-annihilation effect and additional focus on collider signature of doubly charged scalars in a gauged $\rm B-L$ scenarios.}. In these earlier studies, the relic abundance of DM is usually determined by the annihilation cross-section through the freeze-out mechanism, which results in satisfying the correct relic density near the resonances.Here we study the effect of both annihilation 
	and co-annihilations among the dark sector particles in the presence of additional scalar and its implication on the viable parameter space consistent with all phenomenological and experimental constraints.

	The origin of neutrino mass and DM is hitherto not known. Any connection between them is also not established yet. However, it will really be interesting if neutrino mass and the DM phenomenology have an interconnection between them. In light of this, it is worth mentioning here that the spontaneous 
	breaking of $U(1)_{\rm B-L}$ gauge symmetry via the vev of $\Phi_{\rm  BL}$ not only generates sub-eV masses of light 
	neutrinos, but also gives rise to co-annihilations among the dark sector fermions in this study.

	The doubly charged scalar in this model offers novel multi-lepton signatures with missing energy and jets which has already been studied in the literature~\cite{Huitu:1996su,Chun:2003ej,Perez:2008ha,Padhan:2019jlc,Dev:2018kpa,Barman:2019tuo,Bhattacharya:2018fus}. However, as this doubly charged scalar 
	possesses both SM gauge charges as well as $\rm B-L$ gauge charge in this scenario, we study the effect of the $\rm B-L$ gauge boson on the production probability of such a doubly charged scalar. In particular, for a TeV scale doubly charged scalar, we show that the production cross-section can get enhanced significantly if 
	the presence of $\rm B-L$ gauge boson which has mass in a few TeV range. 
	
	The rest of the paper is organized as follows. In section \ref{model}\,, we describe the proposed model, the neutrino mass generation through a variant of type-II seesaw, the scalar masses and mixing. We then discuss how the particles introduced for anomaly cancellation become viable DM candidate and study the relic density in section \ref{dm}\,. In section~\ref{DetCon}, we studied all the relevant constraints from direct, indirect search of DM on our parameter space as well as scrutinized it with respect to the constraint from colliders. We briefly summarize the collider search strategies of the model in section \ref{collider} and finally conclude in section \ref{conclusion}\,. 
	\section{The Model}
	\label{model}
	\begin{table}[ht]
		\resizebox{\linewidth}{!}{
			\begin{tabular}{|c|c|c|c|}
				\hline \multicolumn{2}{|c}{\bf BSM Fields}&  \multicolumn{1}{|c|}{\bf  $\underbrace{ SU(3)_C \otimes SU(2)_L \otimes U(1)_Y}$ $\otimes~ U(1)_{\rm B-L}$} \\ \hline
				{{ {Dark sector Fermions}}} &  ${\chi}_{e_R},~{\chi}_{\mu_R}$&   ~~~~~~~~1 ~~~~~~~~~~~1~~~~~~~~~~~~0~~~~~~~~~~~~-4~~~~~~~~~~  \\ [0.5em] \cline{2-3}
				&  $\chi_{\tau_R}$& ~~~~~~~~1 ~~~~~~~~~~~1~~~~~~~~~~~~0~~~~~~~~~~~~5~~~~~~~~~~  \\ [0.5em] 
				\hline
				\hline
				Heavy Scalars & $\Delta=\begin{pmatrix} \frac{\delta^{+}}{\sqrt{2}} && \delta^{++} \\ 
					\delta^{0} && -\frac{\delta^{+}}{\sqrt{2}}\end{pmatrix}$ & ~~~~~~~~1 ~~~~~~~~~~~3~~~~~~~~~~~~2~~~~~~~~~~~~0~~~~~~~~~~   \\[0.75em]
				\cline{2-3}
				& $\Phi_{\rm BL}$ & ~~~~~~~~1 ~~~~~~~~~~~1~~~~~~~~~~~~0~~~~~~~~~~~~-1~~~~~~~~~~  \\[0.5em]
				\cline{2-3}
				&  $\Phi_{12}$& ~~~~~~~~1 ~~~~~~~~~~~1~~~~~~~~~~~~0~~~~~~~~~~~~8~~~~~~~~~~  \\ [0.5em]
				
				\hline
				\hline
				Light Scalars 
				&  $\xi=\begin{pmatrix} \frac{\xi^{+}}{\sqrt{2}} && \xi^{++} \\ 
					\xi^{0} && -\frac{\xi^{+}}{\sqrt{2}}  \end{pmatrix}$& ~~~~~~~~1 ~~~~~~~~~~~3~~~~~~~~~~~~2~~~~~~~~~~~~2~~~~~~~~~~  \\[0.75em]
				\cline{2-3}
				&  $\Phi_{3}$& ~~~~~~~~~~1 ~~~~~~~~~~~1~~~~~~~~~~~~0~~~~~~~~~~~~-10~~~~~~~~~~  \\ [0.5em]
				\hline
		\end{tabular}}
		\caption{\footnotesize{Charge assignment of BSM fields under the gauge group $\mathcal{G} \equiv \mathcal{G}_{\rm SM} \otimes U(1)_{B-L}$, where $\mathcal{G}_{\rm SM}\equiv SU(3)_C \otimes SU(2)_L \otimes U(1)_Y$. }}
		\label{tab1}
	\end{table}
	
	The model under consideration is  a very well motivated BSM framework based on the gauged $U(1)_{\rm B-L}$ symmetry~\cite{Davidson:1978pm,Mohapatra:1980qe, 
		Marshak:1979fm, Masiero:1982fi, Mohapatra:1982xz, Buchmuller:1991ce} in which we implement a modified type-II seesaw to explain the sub-eV neutrino mass by introducing two triplet scalars $\Delta$ and $\xi$. 
	$\Delta$ is super heavy with $M_{\Delta} \sim 10^{3}$ TeV and $M_{\xi} \sim$ TeV $<<$ $M_{\Delta}$ and the $\rm B-L$ charges of $\Delta$ and $\xi$ are 0 and 2 
	respectively. As already discussed in the previous section, the additional $U(1)_{\rm B-L}$ gauge symmetry introduces $\rm B-L$ anomalies in the theory. To cancel 
	these $\rm B-L$ anomalies we introduce three right chiral dark sector fermions $\chi_{i_R}$ $(i=e,\mu,\tau)$, where the $\rm B-L$ charges of $\chi_{e_R}$,$\chi_{\mu_R}$ and $\chi_{\tau_R}$ are -4, -4, +5 respectively. Note that such unconventional $\rm B-L$ charge assignment of the $\chi_{i_R}\,(i=e,\mu,\tau)$ forbids their Yukawa couplings with the SM particles. Also three singlet scalars: $\Phi_{\rm BL}$, $\Phi_{12}$ and $\Phi_{3}$ with ${\rm B-L}$ charges  -1, +8, -10 are introduced. As a result of which $\Phi_{12}$ and $\Phi_{3}$ couple to $\chi_{e_R,\mu_R}$ and $\chi_{\tau_R}$ respectively through Yukawa terms and 
	the vevs of $\Phi_{12}$ and $\Phi_{3}$ provides Majorana masses to these dark sector fermions.  The vev of $\Phi_{\rm BL}$ provides a small mixing between 
	$\Delta$ and $\xi$ which plays a crucial role in generating sub-eV masses of neutrinos. This vev $\langle \Phi_{{\rm BL}} \rangle$ is also instrumental in controlling the co-annihilations among the dark sector particles and hence is crucial for DM phenomenology too. As a consequence this establishes an interesting correlation 
	between the neutrino mass and DM.The particle content and their charge assignments are listed in Table.\ref{tab1}.
	
	The Lagrangian involving the BSM fields consistent with the extended symmetry is given by:
	\begin{eqnarray}
		\mathcal{L} &=& \mathcal{L}^{\rm SM +BSM ~Scalar} +\mathcal{L}^{\rm DM}
	\end{eqnarray}
	where 
	\begin{eqnarray}
		\mathcal{L}^{\rm SM+BSM ~Scalar} &\supset&  {\big| D_{\mu} H \big|}^2  +{\big| D_{\mu} \Phi_3 \big|}^2+{\big| D_{\mu} \Phi_{\rm BL} \big|}^2 + {\big| D_{\mu} \Phi_{12} \big|}^2 \nonumber \\
		&& + Tr[(D_\mu \Delta)^\dagger (D^\mu \Delta)]+ Tr[(D_\mu \xi)^\dagger (D^\mu \xi )] \nonumber \\
		&& - Y_{ij}^{\xi}~ \overline{L_{i}^c}\,\, i \tau_{2} \, \xi \, L_{j} \nonumber\\
		&&- \mathrm{V}^{\mathbb L}(H,\xi, \Phi_3) - V^{\mathbb H}(\Delta,\Phi_{\rm BL}, \Phi_{12})-V^{\mathbb{LH}} .
		\label{lagrangian}
	\end{eqnarray}
	
	Here $i,j$ runs over all three lepton generations. In the above Lagrangian, $\mathrm{V}^{ \mathbb{L}}$ is the scalar potential involving scalars in sub-TeV mass range ($H,\xi, \Phi_3$), $\mathrm{V}^{\mathbb{H}}$ stands for scalar potential of the heavy fields ($\Delta,\Phi_{\rm BL}, \Phi_{12}$) and the scalar potential which involves both sub-TeV and super heavy fields is defined by $\mathrm{V}^{\mathbb {LH}}$.
	
	The covariant derivatives for these fields can be written as:
	\bea
	D_\mu H &=& \partial_\mu H + igT^a W^a_\mu H + i g' B_\mu H       \nonumber \\
	D_\mu \Delta &=&  \partial_\mu \Delta + ig[T^a W^a_\mu, \Delta] + i g' Y B_\mu \Delta \nonumber\\
	D_\mu \xi &=&  \partial_\mu \xi + ig[T^a W^a_\mu, \xi] + i g' Y B_\mu \xi + i g_{BL} Y_{BL} (Z_{BL})_{\mu} \xi  \nonumber \\
	D_\mu G &=& \partial_\mu G +  i g_{BL} Y_{BL} (Z_{BL})_{\mu} G   ~~~~~~~~~~~~~~~~~~~~~~~                  {\rm ~~~~~~where} ~G=\{\Phi_3,\Phi_{\rm BL},\Phi_{12}\} \nonumber . 
	\eea
	
	The Lagrangian of the dark sector can be written as:
	\begin{eqnarray}
		\mathcal{L}^{\rm DM} &=& \overline{\chi_{e_R}}\, i \gamma^\mu D_{\mu} \chi_{e_R} + \overline{\chi_{\mu_R}}\, i \gamma^\mu D_{\mu} \chi_{\mu_R}+ \overline{\chi_{\tau_R}} i \gamma^\mu D_{\mu} \chi_{\tau_R} \nonumber \\
		& & + Y_{11}\,\, \Phi_{12}\,\, \overline{(\chi_{e_R})^c}\,\, \chi_{e_R}+ Y_{22}\,\, \Phi_{12}\,\, \overline{(\chi_{\mu_R})^c}\,\, \chi_{\mu_R}+ Y_{12}\,\, \Phi_{12}\,\, \overline{(\chi_{e_R})^c}\,\, \chi_{\mu_R} \nonumber\\
		&&   +Y_{13}~\Phi_{_{\rm BL}} \overline{(\chi_{e_R})^c}~  \chi_{\tau_R}+ Y_{23}~ \Phi_{_{\rm BL}} \overline{(\chi_{\mu_R})^c}~  \chi_{\tau_R} \nonumber \\
		&& + Y_{33}\,\, \Phi_{3}\,\, \overline{(\chi_{\tau_R})^c}~ \chi_{\tau_R}+ {\rm h.c.} 
		\label{lagDM}
	\end{eqnarray}
	
	where
	\begin{equation}
		D_{\mu} \chi= \partial_\mu \chi + i~g_{_{_{\rm BL}}} Y_{_{\rm BL}}  (Z_{_{\rm BL}})_\mu \chi \nonumber\\.
	\end{equation}
	The gauge coupling  associated with $U(1)_{\rm B-L}$ is $g_{_{\rm BL}}$ and $Z_{\rm BL}$ is the corresponding gauge boson. 
	The scalar potentials which are mentioned in the Lagrangian~\ref{lagrangian} can be written as:
	\begin{eqnarray}
		\mathrm{V}^{\mathbb{ L}}(H,\xi,\Phi_3) &=& -{\mu}^2_H H^\dagger H + \lambda_H {(H^\dagger H)^2}  + M^2_{\xi}~ Tr[\xi^\dagger \xi]+ \lambda_{\xi} (Tr[\xi^\dagger \xi])^2 + \lambda'_{\xi} Tr[(\xi^\dagger \xi)^2] \nonumber \\
		&&  + \lambda_{\xi H} {Tr[\xi^\dagger \xi](H^\dagger H)}  + \lambda'_{\xi H} \Big(H^\dagger \xi \xi^\dagger H\Big) - {\mu}^2_{\Phi_3} \Phi^\dagger_3 \Phi_3 + \lambda_{\Phi_3} {(\Phi^\dagger_3 \Phi_3)^2} \nonumber \\
		&& + \lambda_{H \Phi_3} (H^\dagger H)(\Phi^\dagger_3 \Phi_3)  +\lambda_{\xi\Phi_3} (\Phi_3^\dagger \Phi_3)Tr[\xi^\dagger \xi] 
		\label{sclarpot1}	
	\end{eqnarray}
	\begin{eqnarray}
		\mathrm{V}^{\mathbb{ H}}(\Delta,\Phi_{\rm BL}, \Phi_{12}) &=& 
		M^2_{\Delta}~ Tr[\Delta^\dagger \Delta]  + \lambda_{\Delta} (Tr[\Delta^\dagger \Delta])^2 + \lambda^\prime_{\Delta} Tr[(\Delta^\dagger \Delta)^2] \nonumber \\
		&& -  {\mu}^2_{\Phi_{{\rm BL}}} \Phi_{{\rm BL}}^\dagger  \Phi_{{\rm BL}} + \lambda_{\Phi_{{\rm BL}}} (\Phi_{{\rm BL}}^\dagger  \Phi_{{\rm BL}})^2 + \lambda_{\Delta \Phi_{{\rm BL}}} Tr[\Delta^\dagger \Delta] (\Phi^\dagger_{{\rm BL}} \Phi_{{\rm BL}}) \nonumber \\
		&& - {\mu}^2_{\Phi_{12}} \Phi_{12}^\dagger \Phi_{12} + \lambda_{\Phi_{12}} {(\Phi_{12}^\dagger \Phi_{12})^2} + \lambda_{\Delta \Phi_{12}} Tr[\Delta^\dagger \Delta] ({\Phi_{12}}^\dagger \Phi_{12})\nonumber \\
		&& +\lambda_{\Phi_{{\rm BL}}\Phi_{12}} (\Phi_{{\rm BL}}^\dagger  \Phi_{{\rm BL}})({\Phi_{12}}^\dagger \Phi_{12}) + \lambda'_{\Phi_{{\rm BL}}\Phi_{12}} (\Phi_{{\rm BL}}^\dagger  \Phi_{12})({\Phi_{12}}^\dagger \Phi_{\rm BL}) 
		\label{sclarpot2}	
	\end{eqnarray}
	\begin{eqnarray}
		\mathrm{V}^{\mathbb{LH}} 
		&=& \lambda_{\Delta H} {Tr[\Delta^\dagger \Delta](H^\dagger H)} + \lambda^\prime_{\Delta H} \Big(H^\dagger \Delta \Delta^\dagger H\Big)+ [\mu_\Delta \Big(H^T i\sigma^2 \Delta^\dagger H\Big)+ h.c] \nonumber \\
		&& +\lambda_{H \Phi_{{\rm BL}}} (H^\dagger H)(\Phi_{{\rm BL}}^\dagger  \Phi_{{\rm BL}}) + \lambda_{H \Phi_{12}} (H^\dagger H)({\Phi_{12}}^\dagger \Phi_{12}) + \lambda_{\Phi_{{\rm BL}}\Phi_3} (\Phi_{{\rm BL}}^\dagger  \Phi_{{\rm BL}})(\Phi^\dagger_3 \Phi_3) \nonumber\\
		&&+ \lambda'_{\Phi_{{\rm BL}}\Phi_3} (\Phi_{{\rm BL}}^\dagger  \Phi_{3})(\Phi^\dagger_3 \Phi_{\rm BL}) + \lambda_{\Phi_{12} \Phi_3}({\Phi_{12}}^\dagger \Phi_{12})({\Phi}^\dagger_3 \Phi_3) + \lambda'_{\Phi_{12} \Phi_3}({\Phi_{12}}^\dagger \Phi_3)(\Phi^\dagger_3 \Phi_{12})\nonumber\\	
		&&  + \lambda_{\Delta \Phi_3} (\Delta^\dagger \Delta) (\Phi^\dagger_3 \Phi_3) + \lambda_{\xi \Phi_{{\rm BL}}} Tr[\xi^\dagger \xi](\Phi^\dagger_{{\rm BL}} \Phi_{{\rm BL}}) + \lambda_{\xi \Phi_{12}}Tr[\xi^\dagger \xi]({\Phi_{12}}^\dagger \Phi_{12}) \nonumber \\
		&&  + \lambda_{\Delta \xi} Tr[\Delta^\dagger \Delta] Tr[\xi^\dagger \xi] + \lambda'_{\Delta \xi} Tr[\Delta^\dagger \xi] Tr[\xi^\dagger \Delta]  \nonumber \\
		&&+  \lambda_{P} \Phi_{{\rm BL}}^2 Tr[\Delta^\dagger \xi] + \lambda_{Q} (\Phi_{{\rm BL}}^\dagger)^2 \Phi_3 \Phi_{12} + h.c.
		\label{sclarpot3}	
	\end{eqnarray}
	Here it is worth mentioning that the mass squared terms of $\Delta$ and $\xi$ are chosen to be positive so they do not get any spontaneous vev.  Only the neutral components of 
	$H$,$\Phi_{12}$,$\Phi_{\rm  BL}$ and $\Phi_3$ acquire non-zero vevs. However, after electroweak phase transition, $\Delta$ and $\xi$ acquire induced vevs.
	
	For simplicity, we assume a certain mass hierarchy among the scalars. The masses of $H$, $\Phi_3$ and $\xi$ are of similar order in 
	sub-TeV range, while the masses of $\Phi_{\rm  BL}$ and $\Phi_{12}$ are in Several TeV scale. 
	To make the analysis simpler, we decouple the light scalar sector from the heavy scalar sector by considering all quartic couplings in the scalar potential $\mathrm{V}^{\mathbb{LH}}$ to be negligible. It is worth mentioning that this assumption does not affect our DM phenomenology. 
	
	We parameterize the low energy neutral scalars as:
	\begin{equation*}
		H^{0}= \frac{v_H + h_H + i\, G_H}{\sqrt{2}}~~,
		~~~~
		\xi^0=\frac{v_{\xi}+ h_\xi + i G_\xi}{\sqrt{2}}~~,
		~~~~
		\Phi_{3}=\frac{v_3+ h_3 + i\, G_{\Phi_3}}{\sqrt{2}} ~~.
	\end{equation*}
	\subsection{NEUTRINO MASS}\label{neutrinomass}
	\begin{figure}[htb!]
		\begin{center}
			\includegraphics[scale=0.35]{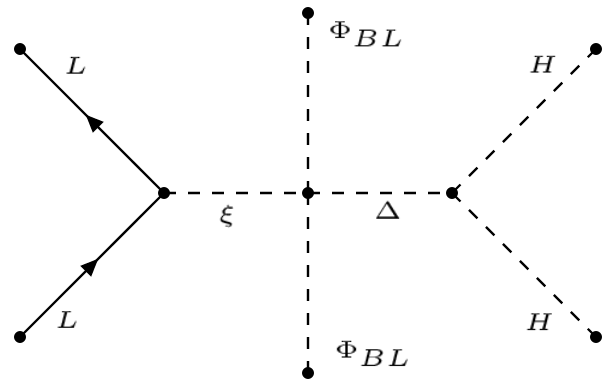}
			\caption{\footnotesize{Generation of neutrino mass through the modified Type-II seesaw.}}
			\label{numass}
		\end{center}
	\end{figure}
	
	The relevant Feynman diagram of this modified type-II seesaw which gives rise to light neutrino masses is shown in Fig~\ref{numass}. In this modified 
	version of type-II seesaw, the conventional heavy triplet scalar $\Delta$ can not generate Majorana masses for light neutrinos as the $\rm B-L$ quantum 
	number of $\Delta$ is zero. However this super heavy scalar $\Delta$ can mix with the TeV scalar triplet $\xi$, once $\Phi_{{\rm BL}}$ acquires vev and 
	breaks the $U(1)_{\rm B-L}$ symmetry spontaneously. By the virtue of the trilinear term of $\Delta$ with SM Higgs doublet $H$, it gets an induced vev after 
	electroweak symmetry breaking similar to the case of traditional type-II seesaw.  The induced vev acquired by $\Delta$ after EW phase transition is given by
	\begin{equation}
		\langle \Delta \rangle = v_{_\Delta} \simeq - \frac{\mu_\Delta v^2_H}{2{\sqrt{2}} M^2_\Delta}\,.
	\end{equation}
	
	Since $\xi$ mixes with $\Delta$ after $U(1)_{\rm B-L}$ breaking, 
	it also acquires an induced vev after EW symmetry breaking which is given by
	\begin{equation}
		\langle \xi \rangle = v_{_\xi} =  -\frac{
			\lambda_{P} v^2_{_{\rm BL}}}{4 M^2_\xi } v_\Delta\,.
	\end{equation}
	Assuming $
	\lambda_{P} v^2_{{\rm BL}} \sim M^2_\xi$, we obtain $v_\Delta \simeq v_\xi$, even if $\xi$ and $\Delta$  have several orders of magnitude difference in their masses.  
	
	As we know that, in the Standard Model the custodial symmetry ensures that the $\rho$ parameter $\rho \equiv \frac{M_W^2}{M_Z^2 \cos^2 \theta}$ is equal to 1 at tree level. However, in the present scenario, because of the presence of the triplet scalars, the form of the $\rho$ parameter gets modified and is given by: 
	\begin{equation}
		\rho = \frac{v^2_{H}+2 v^2_{\Delta}+ 2 v^2_{\xi}}{v^2_{H}+4 v^2_{\Delta}+ 4 v^2_{\xi}} \simeq \frac{1+ 4x^2}{1+8x^2}
	\end{equation}
	where $x= v_\Delta/v_H \simeq v_\xi/v_H$. 
	According to the latest updates of the electroweak
	observables fits, the $\rho$ parameter is constrained as $\rho=1.00038\pm0.00020$\cite{ParticleDataGroup:2020ssz}. This implies that, corresponding to the constraints on the $\rho$ parameter, we get an upper bound on the vev of the triplet $\xi$ of order $\{1.650-2.962\}$ GeV.

	
	After integrating out the heavy degrees of freedom in the Feynman diagram given in Fig.\ref{numass}, we get the Majorana mass matrix of the light neutrinos to be
	\begin{equation}
		(M_\nu)_{ij} = Y_{ij}^{\xi} v_\xi =- Y_{ij}^{\xi}~ \frac{\lambda_{P} v^2_{{\rm BL}}}{4 M^2_\xi} v_\Delta .
		\label{numasseqn}
	\end{equation}
	
	As $v_\Delta \simeq v_\xi \sim \mathcal{O}(1)$ GeV, we can get sub-eV neutrino masses by appropriately tuning the Yukawa couplings. Here it is worth noticing 
	that the mixing between the super heavy triplet scalar $\Delta$ and the TeV scale scalar triplet $\xi$ gives rise to the neutrino mass. Essentially this set-up can be thought of in an effective manner. After the $U(1)_{\rm B-L}$ breaking, $\xi$ develops an effective trilinear coupling  with the SM Higgs {\it i.e.}  $\mu_\xi~ \xi^\dagger H H$ where $\mu_\xi$ is given by $ \mu_\xi={\mu_{\Delta} \langle \Phi_{{\rm BL}}\rangle^2}/{M^2_{\Delta}}$. And this effective coupling is similar to the conventional type-II seesaw which leads to the generation of neutrino mass in this scenario.
	
	In Eq.~\ref{numasseqn}, $(M_\nu)_{ij}$ is a complex $3\times 3$ matrix and can be diagonalized by the PMNS matrix~\cite{Valle:2006vb} for which the standard parametrization is given by:
		\begin{equation}
			U=\left(
			\begin{array}{ccc}
				c_{12}c_{13}                                & s_{12}c_{13}                                & s_{13}e^{-i\delta} \\
				-s_{12}c_{23}-c_{12}s_{23}s_{13}e^{i\delta} & c_{12}c_{23}-s_{12}s_{23}s_{13}e^{i\delta}  & s_{23}c_{13}       \\
				s_{12}s_{23}-c_{12}c_{23}s_{13}e^{i\delta}  & -c_{12}s_{23}-s_{12}c_{23}s_{13}e^{i\delta} & c_{23}c_{13}       \\
			\end{array}
			\right) U_{ph}
			\label{UPMNS}   
		\end{equation}
		where  $c_{ij}\equiv\cos\theta_{ij}$, $s_{ij}\equiv\sin\theta_{ij}$ and $\delta$ is the Dirac phase. Here $U_{ph}$ is given by $U_{ph} = Diag(1, e^{i\alpha_1/2}, e^{i\alpha_2/2} )$  with  $\alpha_{1,2}$ are the CP-violating Majorana phases.
	
	From Eq.~\ref{numasseqn}, we can write the couplings $Y^{\xi}_{ij}$ as follows:
		\begin{equation}
			Y^{\xi}_{ij}=\frac{(M_\nu)_{ij}}{v_\xi}=\frac{1}{v_\xi}\big[U.M^{diag}_\nu.U^T\big]_{ij}
			\label{yuk}
	\end{equation} 
	Neutrino oscillation experiments involving solar, atmospheric, accelerator,
		and reactor neutrinos are sensitive to the mass-squared differences and the mixing
		angles, and the value of these parameters in the $3\sigma$ range used in the analysis\cite{ParticleDataGroup:2020ssz} are as follows.
		\begin{eqnarray}
			\Delta m^2_{sol}\equiv m^2_2 -m^2_1
			\in [6.79-8.01]\times 10^{-5} {\rm eV}^2 \nonumber,~~~
			|\Delta m^2_{atm}|\equiv |m^2_3 -m^2_1|
			\in [2.35,2.54]\times 10^{-3} {\rm eV}^2\,~ \\
			\sin^2\theta_{12}\in [0.27,0.35] \,,~~~~
			\sin^2\theta_{23} \in [0.43,0.60] \,,~~~~
			\sin^2\theta_{13}\in [0.019,0.024]\,.~~~~~~~~~~~~
			\label{obs_para}
		\end{eqnarray}
		Since the sign of $\Delta m_{31}^2$ is undetermined, 
		distinct neutrino mass hierarchies are possible.
		The case with $\Delta m^2_{31} >0$ is referred to as
		{\it Normal hierarchy} (NH) where $m_1 < m_2 < m_3$
		and the case with $\Delta m^2_{31} <0$ is known as
		{\it Inverted hierarchy} (IH) where $m_3 <  m_1 < m_2$.  
		Information on the mass of the lightest neutrino
		and the Majorana phases cannot be obtained from neutrino oscillation experiments as the oscillation probabilities are independent 
		of these parameters. Because of the general texture, the Yukawa couplings in Eq.~\ref{yuk} can facilitate charged lepton flavour violating(CLFV) decays and hence are constrained by the non-observation of such LFV processes at various experiments which we discuss in the  subsection~\ref{CLFV}.
	

	\subsection{SCALAR MASSES \& MIXING}
	\label{scalars}
	As already discussed in the section~\ref{model}, the only significant mixing relevant for low energy phenomenological aspects is the mixing between $H$, $\xi$ and $\Phi_3$ since all other mixings are insignificant and can be neglected. In this section we only consider the light scalar sector, $\mathrm{V}^{\mathbb{L}}(H,\xi,\Phi_3)$ which is relevant for low energy phenomenology. 
	
	The minimization conditions for the scalar potential are given by:
	\begin{eqnarray}
		\mu_H^2 &=& \frac{1}{2} \left(\lambda_{H \Phi_3} v_3^2+2 \lambda_H v_H^2+v_\xi^2 (\lambda_{\xi H}+\lambda'_{\xi H})-2 \sqrt{2} \mu_\xi  v_\xi\right) \nonumber\\
		M_\xi^2 &=& \frac{1}{2} \left(-\lambda_{\xi \Phi_3} v_3^2-v_H^2 (\lambda_{\xi H}+\lambda'_{\xi H})+\frac{\sqrt{2} \mu_\xi  v_H^2}{v_\xi}-2 v_\xi^2 (\lambda_\xi+\lambda'_\xi)\right) \nonumber\\
		\mu_{\Phi_3}^2 &=&\frac{1}{2} \left(2 \lambda_{\Phi_3} v_3^2+\lambda_{H \Phi_3} v_H^2+\lambda_{\xi \Phi_3} v_\xi^2\right) .
	\end{eqnarray}
	The neutral CP even scalar mass terms of the Lagrangian can be expressed as: 
	\begin{eqnarray}
		\mathcal{L}_{\rm mass}&=& \frac{1}{2} \begin{pmatrix} h_H && h_{\xi} && h_3 \end{pmatrix}
		\left(
		\begin{array}{ccc}
			2 \lambda_H v_H^2 & v_H v_\xi (\lambda_{\xi H}+\lambda'_{\xi H}) & \lambda_{H \Phi_3} v_3 v_H \\
			v_H v_\xi (\lambda_{\xi H}+\lambda'_{\xi H})  & \frac{\mu_\xi  v_H^2}{\sqrt{2} v_\xi}+2 v_\xi^2 (\lambda_\xi+\lambda'_\xi) &\lambda_{\xi \Phi_3} v_3 v_\xi \\
			\lambda_{H \Phi_3} v_3 v_H &\lambda_{\xi \Phi_3} v_3 v_\xi & 2 \lambda_{ \Phi_3} v_3^2 \\
		\end{array}
		\right)
		\begin{pmatrix} h_H \\ h_{\xi} \\ h_3 \end{pmatrix} \nonumber \\
		&=&  \frac{1}{2} \begin{pmatrix} h_H && h_{\xi} && h_3 \end{pmatrix} 
		\mathcal{M}^{\rm CP ~Even}
		\begin{pmatrix} h_H \\ h_{\xi} \\ h_3 \end{pmatrix} \nonumber \\
		&=&  \frac{1}{2} \begin{pmatrix} H_1 && H_2 && H_3 \end{pmatrix} 
		\Big( \mathcal{O}^T \mathcal{M}^{\rm CP-Even}~~ \mathcal{O} \Big)
		\begin{pmatrix} H_1 \\ H_2 \\ H_3 \end{pmatrix} \nonumber \\
		&=&  \frac{1}{2} \begin{pmatrix} H_1 && H_2 && H_3 \end{pmatrix} 
		\begin{pmatrix} m_{H_1}^2 & 0 & 0 \\ 0 & m_{H_2}^2 & 0 \\ 0 & 0 & m_{H_3}^2  \end{pmatrix} 
		\begin{pmatrix} H_1 \\ H_2 \\ H_3 \end{pmatrix} .
	\end{eqnarray}
	Here $\mathcal{O}$ is the orthogonal matrix which diagonalises the CP-even scalar mass matrix.
	Thus the flavor eigen states and the mass eigen states of these scalars are related by:
	\begin{eqnarray}
		\begin{pmatrix} h_H \\ h_{\xi} \\ h_3 \end{pmatrix} &=&
		\left(
		\begin{array}{ccc}
			c_{12} c_{13} & c_{13} s_{12} & s_{13} \\
			-c_{12} s_{13} s_{23}-c_{23} s_{12} & c_{12} c_{23}-s_{12} s_{13} s_{23} & c_{13} s_{23} \\
			s_{12} s_{23}-c_{12} c_{23} s_{13} & -c_{12} s_{23}-c_{23} s_{12} s_{13} & c_{13} c_{23} \\
		\end{array}
		\right)
		\begin{pmatrix} H_1 \\ H_2 \\ H_3 \end{pmatrix} 
	\end{eqnarray}
	where we abbreviated $\cos\beta_{ij} = c_{ij}$ and $\sin \beta_{ij} = s_{ij}$ , with $\{ij : 12, 13, 23\}.$

	Apart from these three  CP even physical states, $H_1,~H_2 ~{\rm and}~H_3$ with masses $m_{H_1}=125$ GeV (SM like Higgs),$~m_{H_2}$ $~{\rm and}~m_{H_3}$ respectively; the scalar sector has one massive CP odd scalar, $A^0$ of mass $m_{A^0}$, one massive singly charged scalar, $H^\pm$ of mass $m_{H^\pm}$ and one massive doubly charged scalar, $H^{\pm\pm}$ of mass $m_{H^{\pm\pm}}$. The masses of CP odd and charged states are given by: 
	\bea
	m_{A^0}^2 &=& \frac{\mu_{\xi}  \left(v_H^2+4 v_\xi^2\right)}{\sqrt{2} v_\xi} \nonumber \\
	m_{H^\pm}^2 &=& \frac{2 \sqrt{2} \mu_\xi  v_H^2-\lambda'_{\xi H} v_H^2 v_\xi-2 \lambda'_{\xi H} v_\xi^3+4 \sqrt{2} \mu_\xi  v_\xi^2}{4 v_\xi} \nonumber \\
	m_{H^{\pm\pm}}^2 &=& -\frac{\lambda'_{\xi H} v_H^2}{2}+\frac{\mu_\xi  v_H^2}{\sqrt{2} v_\xi}-\lambda'_\xi v_\xi^2 .
	\eea
	\underline{$U(1)_{\rm B-L}$ Gauge Boson mass}:  
	The $Z_{\rm BL}$ boson acquires mass through the vevs of $\Phi_{{\rm BL}}$, $\Phi_{12}$, $\Phi_3$ which are charged under $U(1)_{{\rm B-L}}$ and is given by:
	\begin{equation}
		M_{Z_{\rm BL}}^{2} \simeq g^2_{_{\rm BL}}(v^2_{{\rm BL}}+64v^2_{12}+100v^2_3).
	\end{equation}
	The quartic couplings of scalars are expressed in term of physical masses, vevs and mixing as:
	\begin{eqnarray}
		\lambda_H &=& \frac{c_{13}^2 \left(c_{12}^2 m_{H_1}^2+m_{H_2}^2 s_{12}^2\right)+m_{H_3}^2 s_{13}^2}{2 v_H^2} \nonumber\\ 
		\lambda_{\xi H}&=&\frac{c_{13} s_{13} s_{23} \left(-c_{12}^2 m_{H_1}^2-m_{H_2}^2 s_{12}^2+m_{H_3}^2\right)+c_{12} c_{13} c_{23} s_{12} \left(m_{H_2}^2-m_{H_1}^2\right)}{v_H v_\xi} +\frac{4 m_{H^\pm}^2}{v_H^2+2 v_\xi^2}-\frac{2 m_{A^0}^2}{v_H^2+4 v_\xi^2} \nonumber \\
		\lambda_\xi &=&\frac{s_{23}^2 \left(s_{13}^2 \left(c_{12}^2 m_{H_1}^2+m_{H_2}^2 s_{12}^2\right)+c_{13}^2 m_{H_3}^2\right)+c_{23}^2 \left(c_{12}^2 m_{H_2}^2+m_{H_1}^2 s_{12}^2\right)-4 m_{H^\pm}^2+2 m_{H^{\pm\pm}}^2}{2 v_\xi^2} \nonumber \\
		& &+ \frac{2 c_{12} c_{23} s_{12} s_{13} s_{23} (m_{H_1}-m_{H_2}) (m_{H_1}+m_{H_2})}{2 v_\xi^2} +\frac{4 m_{H^\pm}^2}{v_H^2+2 v_\xi^2}-\frac{2 m_{A^0}^2}{v_H^2+4 v_\xi^2} \nonumber \\
		\lambda'_\xi&=& -\frac{m_{A^0}^2-m_{H^{\pm\pm}}^2-2 m_{H^\pm}^2}{v_\xi^2}+\frac{4 m_{A^0}^2}{v_H^2+4 v_\xi^2}-\frac{4 m_{H^\pm}^2}{v_H^2+2 v_\xi^2}  \nonumber\\
		\lambda'_{\xi H} &=& \frac{4 m_{A^0}^2}{v_H^2+4 v_\xi^2}-\frac{4 m_{H^\pm}^2}{v_H^2+2 v_\xi^2} \nonumber \\
		\mu_\xi &=& \frac{\sqrt{2} m_{A^0}^2 v_\xi}{v_H^2+4 v_\xi^2} \nonumber\\
		\lambda_{\Phi_3}&=&\frac{m_{H_1}^2 (c_{12} c_{23} s_{13}-s_{12} s_{23})^2+m_{H_2}^2 (c_{12} s_{23}+c_{23} s_{12} s_{13})^2+c_{13}^2 c_{23}^2 m_{H_3}^2}{2 v_3^2} \nonumber\\
		\lambda_{H \Phi_3}&=& \frac{c_{13} c_{23} s_{13} \left(-c_{12}^2 m_{H_1}^2-m_{H_2}^2 s_{12}^2+m_{H_3}^2\right)+c_{12} c_{13} s_{12} s_{23} (m_{H_1}-m_{H_2}) (m_{H_1}+m_{H_2})}{v_3 v_H} \nonumber\\
		\lambda_{\xi \Phi_3}&=&\frac{m_{H_1}^2 (c_{12} s_{13} s_{23}+c_{23} s_{12}) (c_{12} c_{23} s_{13}-s_{12} s_{23})-m_{H_2}^2 (c_{12} s_{23}+c_{23} s_{12} s_{13}) (c_{12} c_{23}-s_{12} s_{13} s_{23})}{v_3 v_\xi} \nonumber \\
		&& +\frac{c_{13}^2 c_{23} m_{H_3}^2 s_{23}}{v_3 v_\xi} .
		\label{couplngs}
	\end{eqnarray}
	\subsection*{\underline{Constraints on scalar sector:}}
	As already discussed in Sec.\ref{neutrinomass}, based on the measurement of the $\rho$ parameter $\rho=1.00038\pm0.00020$\cite{ParticleDataGroup:2020ssz}, the triplet $\xi$ vev $v_\xi$ can have an upper bound of order $\{1.650-2.962\}$ GeV.
	Also the mixing angle between the SM Higgs and the triplet scalar is constrained from
	Higgs decay measurement. As obtained by~\cite{Bhattacharya:2017sml}, this mixing angle $\sin \beta_{12}$ is bounded above, in particular,  $\sin \beta_{12} 
	\lesssim 0.05$ to be consistent with experimental observation of $H_{1} \rightarrow W W^{*}$ ~ \cite{Bhattacharya:2017sml,Barman:2019tuo}. There are similar
	bounds on singlet scalar mixing with the SM Higgs boson. Such bounds come from both theoretical and experimental constraints~\cite{Robens:2015gla,Chalons:2016jeu,Adhikari:2020vqo}. 
	The upper bound on singlet scalar-SM Higgs mixing angle $\sin\beta_{13}$ comes form W boson mass correction~\cite{Lopez-Val:2014jva} at NLO. For
	$250 ~{\rm GeV} < m_{H_3} <850 ~{\rm GeV}$,  $\sin\beta_{13}$ is constrained to be $\sin\beta_{13}<0.2-0.3$ where $m_{H_3}$ is the mass of the third physical Higgs.
	
	For our further discussion, we consider the following benchmark points where all the above mentioned constraints are satisfied as well as the quartic couplings mentioned in Eq.~\ref{couplngs} are within the unitarity and perturbativity limit.  
	\begin{eqnarray}
		\{m_{H_2}=331.779,~m_{H_3}=366.784,~m_{A^0}=331.779,~m_{H^\pm}=369.841,~m_{H^{\pm\pm}}=404.343 \nonumber \\ v_\xi= 2.951 ~ {\rm (in~ GeV)};~ s_{12}=0.03,~s_{23}=s_{13}=0.01~\}~~~~~~
		\label{Higgs_values}
	\end{eqnarray}
	
	This parameter choice in Eq.~\ref{Higgs_values} is for definiteness. It is not exhaustive. One can consider another set of parameters in Eq.~\ref{Higgs_values}, without changing any of the consequences in the dark sector that we study here.

		\subsection{\bf Charged Lepton Flavour Violation}
		\label{CLFV}
		Charged lepton flavour violating (CLFV) decay is a
		promising process to study from beyond standard model(BSM) physics point of view. In the SM, such a process occurs at one-loop level and is suppressed by the smallness of neutrino masses, much beyond the current experimental sensitivity. Therefore, any future observation of such LFV decays like $\mu \to 3e$ or $\mu \to e \gamma$ will definitely be a
		signature of new physics beyond the SM. In our model such CLFV decays can occur at tree level mainly mediated via the triplet scalar $H^{\pm\pm}$ and at one-loop level mediated by $H^{\pm\pm}$ and $H^{\pm}$.

		The branching ratio for $\mu \to 3e$ process which can occur at tree level is given by:
		\begin{eqnarray}
			\label{BRmueee}
			Br(\mu \to 3e)
			&=&
			\frac{\ |Y^{\xi}_{\mu e}|^2 |Y^{\xi}_{ee}|^2 }{ 4 G_F^2 m_{H^{\pm\pm}}^4 }\,
			Br(\mu\to e\overline{\nu}\nu)
		\end{eqnarray}
		where Br$(\mu\to e\overline{\nu}\nu)\simeq 100\%$.

\begin{figure}[htb!]
			$$
			\includegraphics[height=5.5cm,width=6cm]{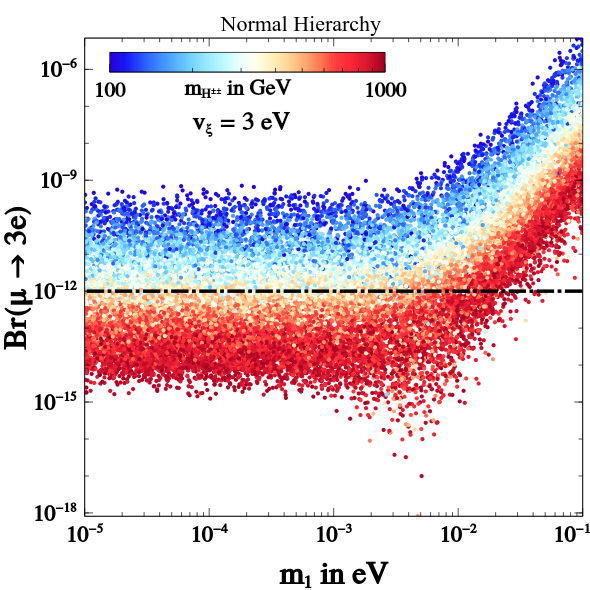}~
			\includegraphics[height=5.5cm,width=6cm]{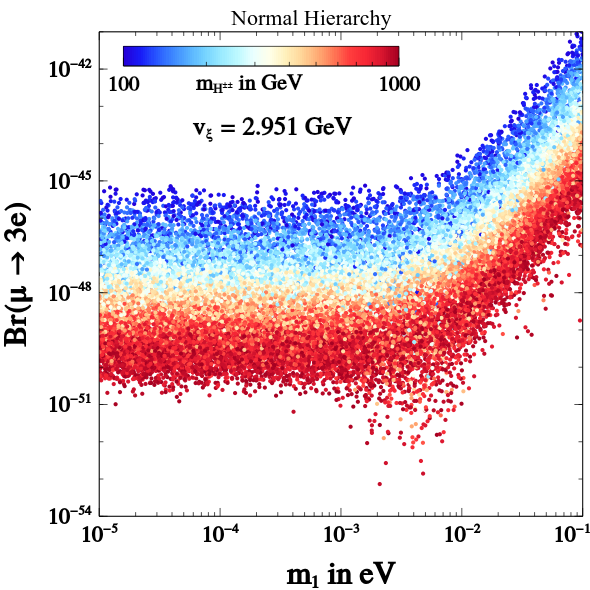}
			$$
			$$
			\includegraphics[height=5.5cm,width=6cm]{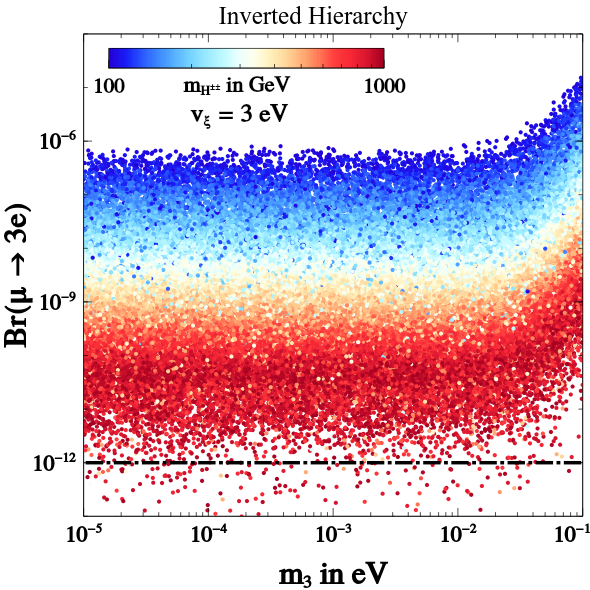}~
			\includegraphics[height=5.5cm,width=6cm]{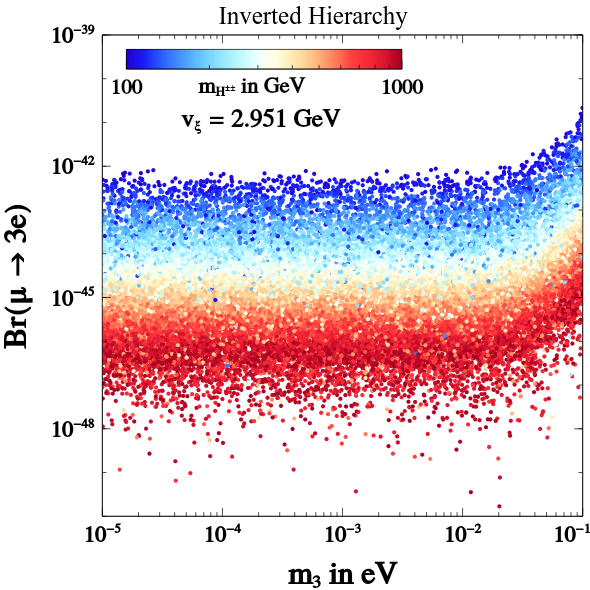}
			$$	
			\caption{$Br(\mu \to 3e)$ as a function of the lightest neutrino mass for both normal and inverted hierarchy. The color code shows the mass of doubly charged scalar $m_{H^{\pm\pm}}$. The black dotted line shows the current bound from~\cite{SINDRUM:1987nra}.}
			\label{muto3eplots}
		\end{figure}

\begin{figure}[htb!]
			$$
			\includegraphics[height=6cm,width=6cm]{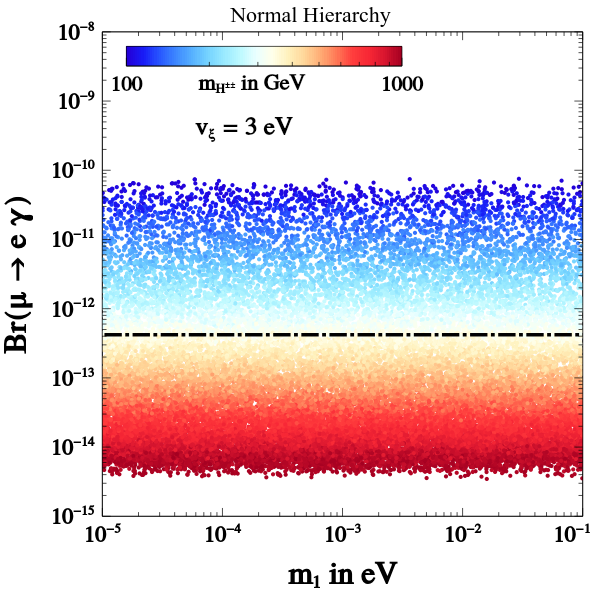}~
			\includegraphics[height=6cm,width=6cm]{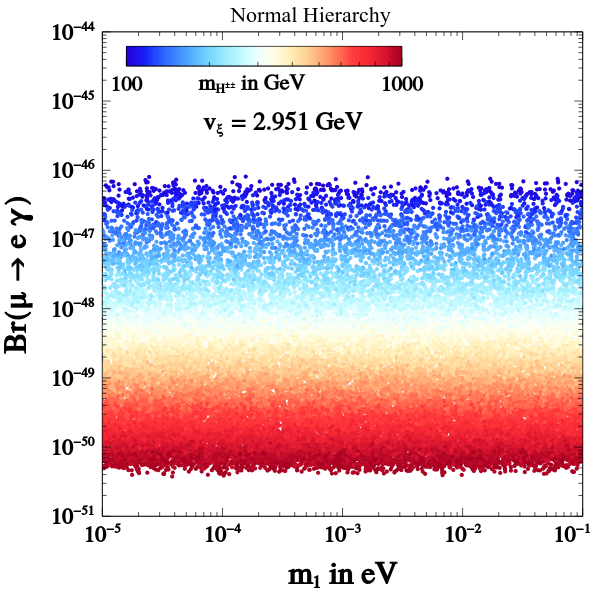}
			$$
			\caption{$Br(\mu \to e \gamma)$ as a function of the lightest neutrino mass for normal hierarchy. The color code shows the mass of doubly charged scalar $m_{H^{\pm\pm}}$. The black dotted line shows the current bound from~\cite{TheMEG:2016wtm}.}
			\label{clfvplots}
	\end{figure}

		Similarly, the branching ratio for $\mu \to e\gamma$ which can take place at loop level is given by (with $m_{H^\pm} \simeq m_{H^{\pm\pm}}$):
		\begin{eqnarray}
			Br(\mu\to e\gamma)
			&\simeq&
			\frac{ 27 \alpha |({Y^{\xi}}^\dagger Y^{\xi})_{e\mu}|^2 }
			{ 64 \pi G_F^2 m_{H^{\pm\pm}}^4 }Br(\mu\to e\overline{\nu}\nu)
		\end{eqnarray}
		We have shown the $Br(\mu \to 3e)$ as a function of the lightest neutrino mass for both normal and inverted hierarchy of neutrino mass spectrum in the upper and lower panel of Fig.~\ref{muto3eplots} and $Br(\mu \to e \gamma)$ has been shown in Fig.~\ref{clfvplots} for only normal hierarchy as $Br(\mu \to e \gamma)$ is not so sensitive to the neutrino mass
		spectrum as pointed out in~\cite{Chun:2003ej,Aoki:2009nu}. The left and right panel figures of Fig.~\ref{muto3eplots} and Fig.~\ref{clfvplots} are for $v_{\xi}=3$ eV and $v_{\xi}=2.951$ GeV respectively. Clearly for $v_{\xi}$ in the eV scale, the constraints from the CLFV can rule out higher values of Yukawa couplings and light $m_{H^{\pm\pm}}$. However if $v_{\xi}$ is in the GeV scale, which is the case for our analysis, ( such that $H^{\pm\pm}$ dominantly decays to $W^{\pm} W^{\pm}$ details of which are given in Appendix~\ref{hppdec}) which is crucial for the collider study of $H^{++}$ in our model discussed in Section.~\ref{collider}, the $Br(\mu \to 3e)$ and $Br(\mu \to e \gamma)$ are far below the present and future senisitivity of these experiments and hence these bounds do not affect our parameter space.	

	\section{Dark Matter}
	\label{dm}
	The $U(1)_{\rm B-L}$ gauge symmetry gets spontaneously broken down by the vev of $\Phi_{12}$, $\Phi_{\rm BL}$ and $\Phi_3$ to a remnant 
	$\mathcal Z_2$ symmetry under which the dark sector fermions: $\chi_{i_R}$ $(i=e,\mu,\tau)$ are assumed to be odd, while all other particles 
	transforms trivially. As a result, the lightest among these fermions becomes a viable candidate of DM and can give rise to the observed relic density by thermal freeze-out mechanism.   
	\subsection{The dark sector fermions and their interactions}
	\label{interactions}
	From Eq.\,\ref{lagrangian}\,, the mass matrix for $\chi_{i_R}$ $(i=e,\mu,\tau)$ in the effective theory can be written as: 
	
	\begin{equation}
		-\mathcal{L}_{\chi}^{mass}= \frac{1}{2}
		\begin{pmatrix}
			\overline{(\chi_{e_R})^c} &
			\overline{(\chi_{\mu_R})^c} & 
			\overline{(\chi_{\tau_R})^c} \\
		\end{pmatrix}
		\mathcal{M}
		\begin{pmatrix}
			\chi_{e_R} \\
			\chi_{\mu_R} \\
			\chi_{\tau_R} \\ 
		\end{pmatrix}
	\end{equation}
	where
	\begin{equation}
		\mathcal{M}= 
		\frac{1}{\sqrt{2}}\begin{pmatrix}
			Y_{11}v_{_{12}} && Y_{12}v_{_{12}} && Y_{13}v_{_{\rm BL}}\\
			Y_{12}v_{_{12}} && Y_{22}v_{_{12}} && Y_{23}v_{_{\rm BL}}\\
			Y_{13}v_{_{\rm BL}} && Y_{23}v_{_{\rm BL}} && Y_{33}v_{3}\\ 
		\end{pmatrix} = 
		\begin{pmatrix}  
			[M_{12}] &&  [M'] \\
			[M']^T   &&  M_3  \\ 
		\end{pmatrix}.
		\label{fermion_mass_matrix}
	\end{equation}
	Here $M_{12}, M', M_3$ are:
	\begin{center}
		$M_{12}= 
		\frac{1}{\sqrt{2}}	\begin{pmatrix}
			Y_{11} v_{_{12}} && Y_{12} v_{_{12}}\\
			Y_{12} v_{_{12}} && Y_{22} v_{_{12}}\\
		\end{pmatrix}$
		, 
		$M'=
		\frac{1}{\sqrt{2}}	\begin{pmatrix} 
			Y_{13}v_{_{\rm BL}}\\
			Y_{23}v_{_{\rm BL}}\\
		\end{pmatrix}$
		, 
		$M_3=
		\frac{1}{\sqrt{2}}	\begin{pmatrix} 
			Y_{33}v_{3}
		\end{pmatrix}$.	
	\end{center}

	To capture the co-annihilation effect in the dark sector in a simplest way, we assume 
	\begin{equation}\label{condition}
		Y_{11}=Y_{22}\,,~~~~ Y_{13}=Y_{23}~~~~{\rm and}~~~~Y_{12}<<1\,.
	\end{equation}
	With this assumption two of the dark sector fermions $\chi_e$ and $\chi_\mu$ become almost degenerate and their mixing with the DM $\chi_\tau$ 
	will be defined by a single mixing angle. Moreover, the mass splitting between the DM and NLSP (next to lightest stable particle) 
	will be unique as we discuss below. However, relaxation of this assumption \ref{condition} will lead to two mass splittings and three mixing angles 
	in the dark sector, which make our analysis unnecessarily complicated without implying any new features. So without loss of generality we assert to Eqn. \ref{condition} in the 
	following analysis.

	Using Eq. \ref{condition}, the above Majorana fermion mass matrix $\mathcal{M}$ can be exactly diagonalized 
	by an orthogonal rotation $\mathcal{R} = \mathcal{R}_{13}(\theta).\mathcal{R}_{23}(\theta_{23}=0).\mathcal{R}_{12}(\theta_{12}=\frac{\pi}{4})$ which is 
	essentially characterized by only one parameter $\theta$~\cite{Bhattacharya:2020wra}. So we diagonalized the mass matrix $\mathcal{M}$ as $\mathcal{R}.\mathcal{M}.\mathcal{R}^{T}= 
	\mathcal{M}_{Diag.}$, where the $\mathcal{R}$ is given by:
	\begin{equation}
		\mathcal{R}= \begin{pmatrix}\frac{1}{\sqrt{2}} \cos \theta~~ &~~ \frac{1}{\sqrt{2}}\cos\theta~~ &~~ \sin\theta \\
			-\frac{1}{\sqrt{2}} & \frac{1}{\sqrt{2}} & 0 \\              -\frac{1}{\sqrt{2}}\sin\theta & -\frac{1}{\sqrt{2}}\sin\theta &~~ \cos\theta\\ 
		\end{pmatrix}  .
	\end{equation}	
	
	The rotation parameter $\theta$ required for the diagonalization is given by:
	\begin{eqnarray}
		\tan2\theta \ & = & \ \frac{2\sqrt{2}~~Y_{13} v_{_{\rm BL}}}{(Y_{11}+Y_{12})v_{12}-Y_{33}v_{3}} .
		\label{dm_mixing_angle}
	\end{eqnarray}
	
	Thus the physical states of the dark sector are $\chi_i=\frac{\chi_{i_R}+(\chi_{i_R})^c}{\sqrt{2}}$ and are related to the flavour eigenstates by the following linear combinations:
	\begin{eqnarray}
		\chi_{1_R} \ & = & \frac{1}{\sqrt{2}}\cos\theta ~\chi_{e_R}+\frac{1}{\sqrt{2}}\cos\theta~ \chi_{\mu_R}+\sin\theta~ \chi_{\tau_R}\nonumber\\	
		\chi_{2_R} \ & = & -\frac{1}{\sqrt{2}}~ \chi_{e_R}+\frac{1}{\sqrt{2}} ~\chi_{\mu_R}\nonumber\\	
		\chi_{3_R} \ & = & -\frac{1}{\sqrt{2}}\sin\theta ~\chi_{e_R}-\frac{1}{\sqrt{2}}\sin\theta ~\chi_{\mu_R}+\cos\theta ~\chi_{\tau_R}	.	
	\end{eqnarray}
	
	And the corresponding mass eigenvalues are given by:
	\begin{eqnarray}
		M_1 \ & = & \ \frac{1}{2\sqrt{2}} \bigg[(Y_{11}+Y_{12})v_{12}+Y_{33}v_3+\sqrt{((Y_{11}+Y_{12})v_{12}-Y_{33}v_3)^2+8(Y_{13}v_{_{\rm BL}})^2}\bigg]\nonumber\\
		M_2 \ & = & \ \frac{1}{\sqrt{2}}(Y_{11}-Y_{12})v_{12}\nonumber\\
		M_3 \ & = & \ \frac{1}{2\sqrt{2}} \bigg[(Y_{11}+Y_{12})v_{12}+Y_{33}v_3-\sqrt{((Y_{11}+Y_{12})v_{12}-Y_{33}v_3)^2+8(Y_{13}v_{_{\rm BL}})^2}\bigg].
	\end{eqnarray}
	
	Here it is worthy to mention that in the limit of $Y_{13} \rightarrow 0$ {\it i.e.} $\theta \rightarrow 0$, we get the mass eigen values of the DM particles as $M_{1,2}=\frac{1}{\sqrt{2}}(Y_{11}\pm Y_{12}) v_{12}$ and $M_3 = \frac{1}{\sqrt{2}}Y_{33} v_3$ and the corresponding mass eigen states are $\chi_{1_R,2_R} = \frac{1}{\sqrt{2}}(\chi_{\mu_R} \pm \chi_{e_R})$ and $\chi_{3_R} = \chi_{\tau_R}$. If we assume that the off diagonal Yukawa couplings:  {\it i.e.} $Y_{12},Y_{13} << 1$, then $\chi_{_1}$ and $\chi_{_2}$ become almost degenerate (i.e $M_1 \simeq M_2$). 
	
	We assume $\chi_3$ to be the lightest state which represents the DM candidate, while $\chi_1$ and $\chi_{_2}$ are NLSPs which are almost degenerate. 
	Using the relation $\mathcal{R}.\mathcal{M}.\mathcal{R}^{T}= \mathcal{M}_{Diag.}$, one can express the following relevant parameters in terms of the
	physical masses $M_1,M_3$ and the mixing angle $\sin\theta$ as
	\begin{eqnarray}
		v_{3} &=& \frac{\sqrt{2}}{Y_{33}}\Big(M_{1} \sin^{2}\theta+M_{3} \cos^{2}\theta\Big) \nonumber\\
		v_{12} &=& \frac{\sqrt{2}}{Y_{33}+Y_{12}}\Big(M_{1} \cos^{2}\theta+M_{3} \sin^{2}\theta\Big) \nonumber\\
		Y_{13} &=& \frac{\Delta M \sin2\theta}{2v_{\rm BL}}, 
		\label{dependent_1}
	\end{eqnarray}
	where $\Delta M$ is the mass splitting between the DM and NLSPs i.e. $\Delta M = M_1 - M_3 $.
	
	The gauge coupling $g_{\rm BL}$ can be expressed as 
	\begin{eqnarray}
		g_{_{\rm BL}}&\simeq& \frac{M_{Z_{\rm BL}}}{\sqrt{\big(v^2_{\rm BL}+64~v^2_{12}+100 v_3^2 \big)}}.
		\label{gbl}
	\end{eqnarray}
	
	The flavour eigenstates can be expressed in terms of the physical eigenstates as follows:
	\begin{eqnarray}
		\chi_{e_R} & = & \frac{1}{\sqrt{2}}\cos\theta~\chi_{1_R}-\frac{1}{\sqrt{2}}~\chi_{2_R}-\frac{1}{\sqrt{2}}\sin\theta~\chi_{3_R}\nonumber\\
		\chi_{\mu_R} & = & \frac{1}{\sqrt{2}}\cos\theta~\chi_{1_R}+\frac{1}{\sqrt{2}}~\chi_{2_R}-\frac{1}{\sqrt{2}}\sin\theta~\chi_{3_R}\nonumber\\
		\chi_{\tau_R} & = & \sin\theta~\chi_{1_R}+\cos\theta~\chi_{3_R}  .
	\end{eqnarray}
	\subsection*{DM Interactions}
	The Yukawa and gauge interactions of DM relevant for the calculation of relic density can be written in the physical eigen states as follows:
	\begin{eqnarray}
		\mathcal{L}_{\rm Yuk.} & = & Y_{33} h_3 \overline{(\chi_{\tau_R})^c} \chi_{\tau_R}\nonumber\\
		&=&Y_{33}\bigg[ (s_{12} s_{23}-c_{12} c_{23} s_{13})H_1 -( c_{12} s_{23}+c_{23} s_{12} s_{13})H_2+ ( c_{13} c_{23})H_3\bigg] \overline{(\chi_{\tau_R})^c} \chi_{\tau_R} \nonumber\\
		&=& Y_{33}\bigg[ (s_{12} s_{23}-c_{12} c_{23} s_{13})H_1 -( c_{12} s_{23}+c_{23} s_{12} s_{13})H_2+ ( c_{13} c_{23})H_3\bigg]\nonumber\\ &\times&\bigg[\sin^2\theta~\overline{(\chi_{1_R})^c} \chi_{1_R}+\cos^2\theta~\overline{(\chi_{3_R})^c} \chi_{3_R} + \sin\theta\cos\theta~\big(\overline{(\chi_{1_R})^c} \chi_{3_R}+\overline{(\chi_{3_R})^c} \chi_{1_R}\big)\bigg]\nonumber\\
		\label{interaction_1}
	\end{eqnarray} 
	and 
	\begin{eqnarray}
		\mathcal{L}_{Z_{\rm BL}} \ &  =  & \ g_{_{\rm BL}}\bigg[(4\cos^2\theta+5\sin^2\theta) ~\overline{\chi_{1_R}} ~\gamma^\mu ~\chi_{1_R} + 4 ~\overline{\chi_{2_R}}~ \gamma^\mu~ \chi_{2_R} \nonumber\\&+& (4\sin^2\theta+5\cos^2\theta)~ \overline{\chi_{3_R}} ~\gamma^\mu ~\chi_{3_R}+\cos\theta\sin\theta~(\overline{\chi_{1_R}} ~\gamma^\mu ~\chi_{3_R} + ~\overline{\chi_{3_R}} ~\gamma^\mu~ \chi_{1_R})\bigg] (Z_{\rm BL})_\mu. \nonumber \\
		\label{interaction_2}
	\end{eqnarray}
	
	Note that there is no co-annihilation of $\chi_{3}$ with $\chi_2$. 
	The dominant annihilation and co-annihilation channels for DM are shown in Fig~{\ref{Feyn_diag1}-\ref{Feyn_diag4}}.			
	\begin{figure}[htb!]
		\begin{center}
			\begin{tikzpicture}[line width=0.6 pt, scale=1.2]
				\draw[solid] (2.4,1)--(3.4,0);
				\draw[solid] (2.4,-1)--(3.4,0);
				\draw[dashed] (3.4,0)--(4.4,0);
				\draw[solid] (4.4,0)--(5.4,1);
				\draw[solid] (4.4,0)--(5.4,-1);
				\node  at (2.1,-1) {$\chi_{3},\chi_1$};
				\node at (2.1,1) {$\chi_{3}$};
				\node [above] at (3.9,0.05) {$H_{1,2,3}$};
				\node at (6.4,1.0){$f$ $(= \ell,q)$};
				\node at (5.6,-1.0) {$\overline{f}$};
				\draw[solid] (8.4,1)--(9.4,0);
				\draw[solid] (8.4,-1)--(9.4,0);
				\draw[snake] (9.4,0)--(10.4,0);
				\draw[solid] (10.4,0)--(11.4,1);
				\draw[solid] (10.4,0)--(11.4,-1);
				\node  at (8.0,-1) {$\chi_{3},\chi_1$};
				\node at (8.0,1) {$\chi_3$};
				\node [above] at (9.9,0.05) {$Z_{\rm BL}$};
				\node at (12.5,1.0){$f$ $(= \ell, \nu_{\ell},q)$};
				\node at (12,-1.0) {$\overline{f}$};
			\end{tikzpicture}
		\end{center}
		\caption{Feynman diagrams for DM Annihilation and Co-annihilations: $\chi_3~\chi_{3,1} \to f \overline{f}$.}
		\label{Feyn_diag1}
	\end{figure}
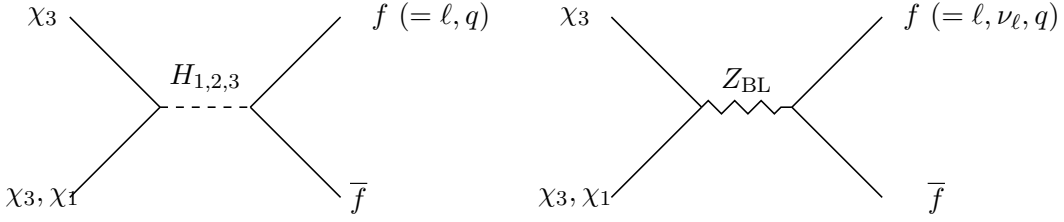
	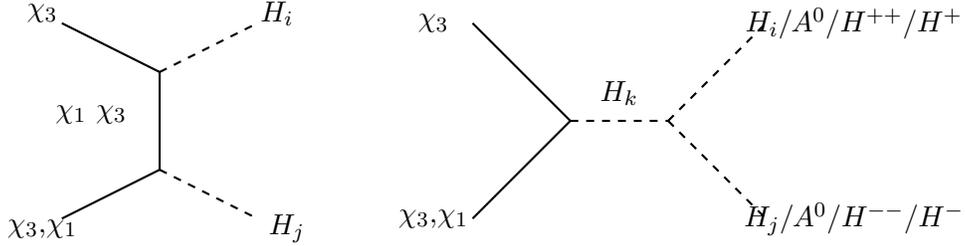
\begin{figure}[htb!]
		\begin{center}
			\begin{tikzpicture}[line width=0.8 pt, scale=1.3]
				\draw[solid] (-1.8,1.0)--(-0.8,0.5);
				\draw[solid] (-1.8,-1.0)--(-0.8,-0.5);
				\draw[solid] (-0.8,0.5)--(-0.8,-0.5);
				\draw[dashed] (-0.8,0.5)--(0.2,1.0);
				\draw[dashed] (-0.8,-0.5)--(0.2,-1.0);
				\node at (-2.0,1.1) {$\chi_3$};
				\node at (-2.0,-1.1) {$\chi_3$,$\chi_1$};
				\node at (-1.5,0.07) {$\chi_1$ $\chi_3$};
				\node at (0.4,1.1) {$H_i$};
				\node at (0.5,-1.1) {$H_{j}$};
				%
				\draw[solid] (2.4,1)--(3.4,0);
				\draw[solid] (2.4,-1)--(3.4,0);
				\draw[dashed] (3.4,0)--(4.4,0);
				\draw[dashed] (4.4,0)--(5.4,1);
				\draw[dashed] (4.4,0)--(5.4,-1);
				\node  at (2.0,-1) {$\chi_3$,$\chi_1$};
				\node at (2.0,1) {$\chi_3$};
				\node [above] at (3.9,0.05) {$H_{k}$};
				\node at (6.3,1.0){$H_i/A^0/H^{++}/H^{+}$};
				\node at (6.3,-1.0) {$H_j/A^0/H^{--}/H^{-}$};
			\end{tikzpicture}
		\end{center}
		\caption{Feynman diagrams for DM Annihilation and Co-annihilations: $\chi_3~\chi_{3,1}\to H_i~H_j$, $A^0~A^0$ $H^+~H^-$, $H^{++}~H^{--}$ ($i,j,k=1,2,3$).}
		\label{Feyn_diag2}
	\end{figure}
	\begin{figure}[htb!]
		\begin{center}
			\begin{tikzpicture}[line width=0.6 pt, scale=1.3]
				\draw[solid] (-1.8,1.0)--(-0.8,0.5);
				\draw[solid] (-1.8,-1.0)--(-0.8,-0.5);
				\draw[solid] (-0.8,0.5)--(-0.8,-0.5);
				\draw[dashed] (-0.8,0.5)--(0.2,1.0);
				\draw[snake] (-0.8,-0.5)--(0.2,-1.0);
				\node at (-2.0,1.1) {$\chi_3$};
				\node at (-2.0,-1.1) {$\chi_3$,$\chi_1$};
				\node at (-1.5,0.07) {$\chi_1,$ $\chi_3$};
				\node at (0.4,1.1) {$H_i$};
				\node at (0.5,-1.1) {$Z_{\rm BL}$};
				%
				\draw[solid] (2.4,1)--(3.4,0);
				\draw[solid] (2.4,-1)--(3.4,0);
				\draw[snake] (3.4,0)--(4.4,0);
				\draw[dashed] (4.4,0)--(5.4,1);
				\draw[snake] (4.4,0)--(5.4,-1);
				\node  at (2.0,-1) {$\chi_3$,$\chi_1$};
				\node at (2.0,1) {$\chi_3$};
				\node [above] at (3.9,0.05) {$Z_{\rm BL}$};
				\node at (5.8,1.0){$H_i$};
				\node at (5.8,-1.0) {$Z_{\rm BL}$};
			\end{tikzpicture}
		\end{center}
		\caption{Feynman diagrams for DM Annihilation and Co-annihilation: $\chi_3~\chi_{3,1} \to H_i~Z_{BL}$ ($i=1,2,3$).}
		\label{Feyn_diag3}
	\end{figure}
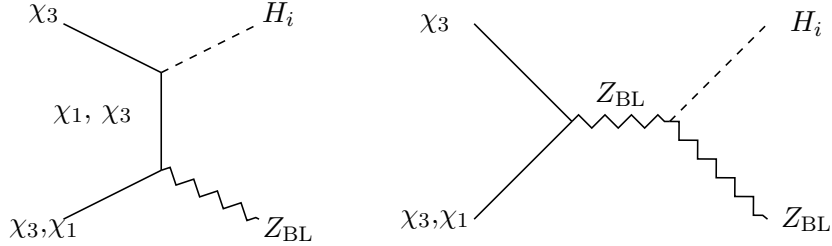
	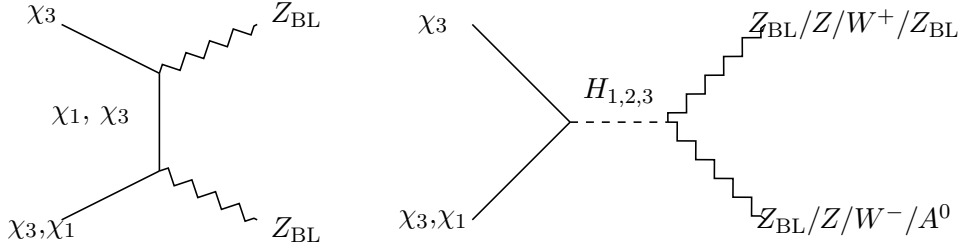
\begin{figure}[htb!]
		\begin{center}
			\begin{tikzpicture}[line width=0.6 pt, scale=1.3]
				\draw[solid] (-1.8,1.0)--(-0.8,0.5);
				\draw[solid] (-1.8,-1.0)--(-0.8,-0.5);
				\draw[solid] (-0.8,0.5)--(-0.8,-0.5);
				\draw[snake] (-0.8,0.5)--(0.2,1.0);
				\draw[snake] (-0.8,-0.5)--(0.2,-1.0);
				\node at (-2.0,1.1) {$\chi_3$};
				\node at (-2.0,-1.1) {$\chi_3$,$\chi_1$};
				\node at (-1.5,0.07) {$\chi_1,$ $\chi_3$};
				\node at (0.6,1.1) {$Z_{\rm BL}$};
				\node at (0.6,-1.1) {$Z_{\rm BL}$};
				%
				\draw[solid] (2.4,1)--(3.4,0);
				\draw[solid] (2.4,-1)--(3.4,0);
				\draw[dashed] (3.4,0)--(4.4,0);
				\draw[snake] (4.4,0)--(5.4,1);
				\draw[snake] (4.4,0)--(5.4,-1);
				\node  at (2.0,-1) {$\chi_3$,$\chi_1$};
				\node at (2.0,1) {$\chi_3$};
				\node [above] at (3.9,0.05) {$H_{1,2,3}$};
				\node at (6.3,1.0){$Z_{\rm BL}/Z/W^+$$/Z_{\rm BL}$};
				\node at (6.3,-1.0) {$Z_{\rm BL}/Z/W^-$$/A^0$};
			\end{tikzpicture}
		\end{center}
		\caption{Feynman diagrams for DM Annihilation and Co-annihilation: $\chi_3~\chi_{3,1} \to Z_{\rm BL}~Z_{\rm BL};~ Z~Z ; ~ W^+~W^-$ and $Z_{\rm BL}~ A^0$} .
		\label{Feyn_diag4}
	\end{figure}				
	\subsection{Relic abundance of DM}
\begin{figure}[h!]
		$$
		\includegraphics[scale=0.4]{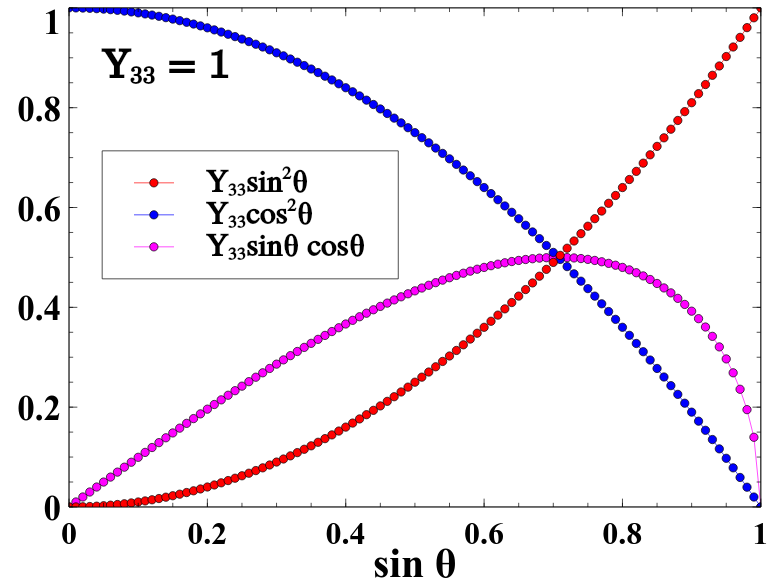}~~
		\includegraphics[scale=0.395]{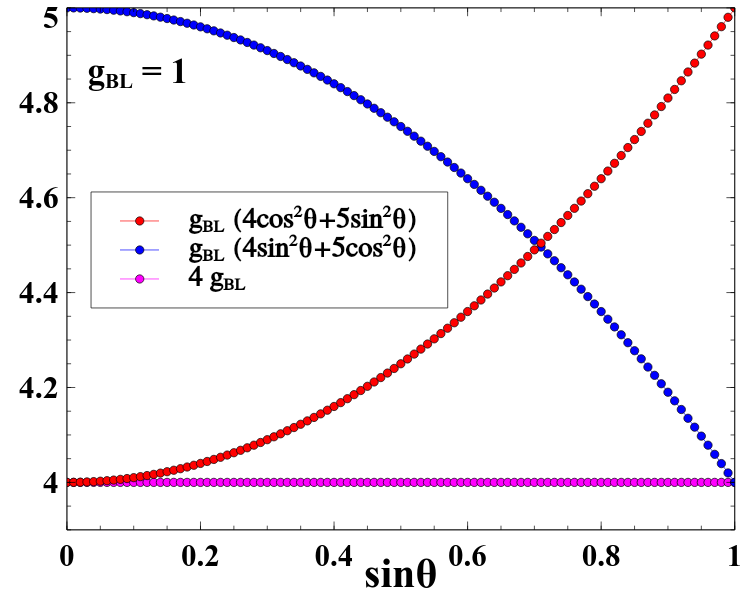}
		$$
		\caption{[Left]: The effective couplings in Yukawa interactions given in Eq.~\ref{interaction_1}. [Right] The effective couplings 
			in gauge interactions given in Eq.~\ref{interaction_2}. }
		\label{coupling}
	\end{figure}

	The DM phenomenology is mainly governed by the following independent parameters:
	
	\begin{eqnarray}
		\{~M_3 \equiv M_{DM},~~\Delta M \equiv (M_1-M_3)\simeq (M_2-M_3), ~\sin\theta, ~ Y_{33}~,~M_{Z_{\rm BL}}\}.
	\end{eqnarray}
	while the other independent parameters that are kept fixed are: $v_{\rm BL}=10$ TeV ,~~$Y_{12}=10^{-6}$, and the dependent parameters are $g_{\rm BL}, v_3, v_{12}$ and $Y_{13}$ as mentioned in Eq~\ref{dependent_1},\ref{gbl}. Depending on the relative magnitudes of these parameters, DM relic can be generated dominantly by annihilation or co-annihilation or a combination of both. The variation of the effective couplings, involved in the annihilation and co-annihilation processes of DM as given in Eq.~\ref{interaction_1} and Eq.~\ref{interaction_2}, with the dark fermions mixing angle $\sin\theta$, which plays a crucial role in DM phenomenology, can be visualized as shown in Fig~\ref{coupling}. Clearly in the limit $\sin\theta\to 0$, the Yukawa coupling involved DM annihilation processes ($ \propto Y_{33} \cos^2 \theta$) dominate, whereas for $\sin\theta \to 1$, the Yukawa coupling involved co-annihilation processes ($\propto Y_{33} \sin^2\theta, Y_{33} \sin \theta \cos \theta $) dominate and play a crucial role in determining the correct relic density. The gauge coupling involved annihilation and co-annihilation processes are almost comparable irrespective of the values of $\sin \theta$.
	
	\vspace{1cm}
	The relic density of DM in this scenario can be estimated by solving the Boltzmann equation in the following form:
	\begin{equation}
		\frac{dn}{dt} + 3Hn = -\langle\sigma v\rangle_{eff}\Big(n^2 - n^2_{eq}\Big),
	\end{equation}
	where $n$ denotes the relic of dark sector fermions, {\it i.e.} $n=n_{\chi_1}+n_{\chi_2}+n_{\chi_3}$ and $n_{eq}= g~({M_{DM}T}/{2\pi})^{3/2}\exp(-M_{DM}/T)$ is
	equilibrium distribution. Here $g=g_1+g_2+g_3$ where $g_3$, $g_1$, $g_2$ are the internal  degrees of freedom of $\chi_{_3}$, $\chi_{_1}$ and $\chi_{_2}$ respectively.  The DM freezes out giving us the thermal relic depending on $\langle\sigma v\rangle_{eff}$, 
		which takes into account all number changing process listed in Fig~{\ref{Feyn_diag1}-\ref{Feyn_diag4}}.	
		This can be written as:
		\begin{align}
			\langle\sigma v\rangle_{eff} &= \frac{g^2_3}{g^2_{eff}}\langle\sigma v\rangle_{\chi_{_3}\chi_{_3}}+ \frac{2g_3 g_1}{g^2_{eff}}\langle\sigma v\rangle_{\chi_{_3}\chi_{_1}}\Big(1+\frac{\Delta M}{M_{DM}}\Big)^{\frac{3}{2}}\exp(-x \frac{\Delta M}{M_{DM}})\nonumber\\
			& + \frac{g^2_1}{g^2_{eff}}\langle\sigma v\rangle_{\chi_{_1}\chi_{_1}}\Big(1+\frac{\Delta M}{M_{DM}}\Big)^{3}\exp(-2x \frac{\Delta M}{M_{DM}}) \nonumber \\
			& + \frac{g^2_2}{g^2_{eff}}\langle\sigma v\rangle_{\chi_{_2}\chi_{_2}}\Big(1+\frac{\Delta M}{M_{DM}}\Big)^{3}\exp(-2x \frac{\Delta M}{M_{DM}}).
			\label{sigmaveff}
		\end{align}
		Here $g_{eff}$ is the effective degrees of freedom which can be expressed as,
		\begin{equation}
			\begin{aligned}
				g_{eff} &= g_3 + g_1 \Big(1+\frac{\Delta M}{M_{DM}}\Big)^{\frac{3}{2}}\exp(-x \frac{\Delta M}{M_{DM}}) + g_2 \Big(1+\frac{\Delta M}{M_{DM}}\Big)^{\frac{3}{2}}\exp(-x \frac{\Delta M}{M_{DM}}) 
			\end{aligned}
			\label{geff}
	\end{equation}
	and the dimensionless parameter $x$ is defined as $x=\frac{M_{DM}}{T}=\frac{M_3}{T}$.	
	
	Eq.~\ref{sigmaveff} can be written in a precise form for convenience in discussion as:
	\begin{equation}
		\begin{aligned}
			\langle\sigma v\rangle_{eff} &= \frac{g^2_3}{g^2_{eff}}\langle\sigma v\rangle_{\chi_{_3}\chi_{_3}}+ \langle\sigma v\rangle_{\chi_{_3}\chi_{_1}}f(\Delta M,M_{DM})\\&+ \langle\sigma v\rangle_{\chi_{_1}\chi_{_1}}h_1(\Delta M,M_{DM}) + \langle\sigma v\rangle_{\chi_{_2}\chi_{_2}}h_2(\Delta M,M_{DM})
		\end{aligned}\\
		\label{sigmaveff2}
	\end{equation} 
	
	where $f,h_1$ and $ h_2$ are the  factors multiplied to the co-annihilation cross-sections which are functions of $\Delta M$ and $M_{\rm DM}$. 
	

	The relic density of the DM ($\chi_{_3}$) then 
	can be given by {\cite{Griest:1990kh,Chatterjee:2014vua,Patra:2014sua}:
		\begin{equation}
			\Omega_{\chi_{_3}}h^2 = \frac{1.09\times 10^9 {\rm GeV}^{-1}}{g^{1/2}_* M_{Pl}}\frac{1}{J(x_f)}
		\end{equation}
		where $g_*=106.7$ and $J(x_f)$ is given by 
		\begin{equation}
			J(x_f)=\int_{x_f}^{\infty} \frac{\langle\sigma v\rangle_{eff}}{x^2} dx  ~~ .
		\end{equation}
		Here $x_f = \frac{M_{DM}}{T_f}$, and $T_f$ denotes the freeze-out temperature of the DM $\chi_{_3}$. We may note here that 
		for correct relic $x_f \simeq 25$.
		
		It is worth mentioning here that we used the package {\tt MicrOmegas}~\cite{Belanger:2008sj} for computing annihilation cross-section and relic 
		density, after generating the model files using {\tt LanHEP}~\cite{Semenov:2014rea}.
		\subsection{Parameter space scan}
		To understand the DM relic density and the specific role of the model parameters in giving rise to the observed relic density, we performed several analyses and scan for allowed parameter space. As discussed in section~\ref{interactions}, the important relevant parameters controlling the relic abundance of DM are: the mass of DM ($M_{DM}$), mass splitting($\Delta M$) between the DM ($\chi_{_3}$) and the next to lightest stable particle ($\chi_{_2}$ and $\chi_{_1}$ as $M_1 \simeq M_2$), and the mixing angle $\sin \theta$. Apart from these three, another crucial parameter that has a noteworthy effect on DM relic, as well as other phenomenological aspects, 
		is the Yukawa coupling $Y_{33}$. We also keep the $\rm B-L$ gauge boson mass ($M_{Z_{\rm BL}}$) as a free parameter. The dependent parameters have already been mentioned in Eqs.~\ref{dependent_1} and ~\ref{gbl}. The other parameters that are kept fixed judiciously during the analysis are $v_{\rm BL}=10$ TeV and 
		$Y_{12}=10^{-6}$. The masses and mixing of Higgses are fixed as per Eq.~\ref{Higgs_values}. 
		\begin{figure}[htb!]
			$$
			\includegraphics[scale=0.26]{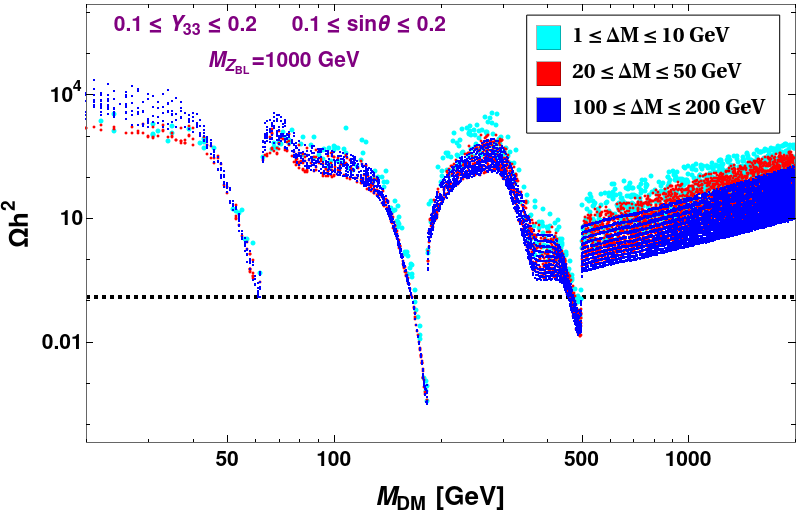}~
			\includegraphics[scale=0.26]{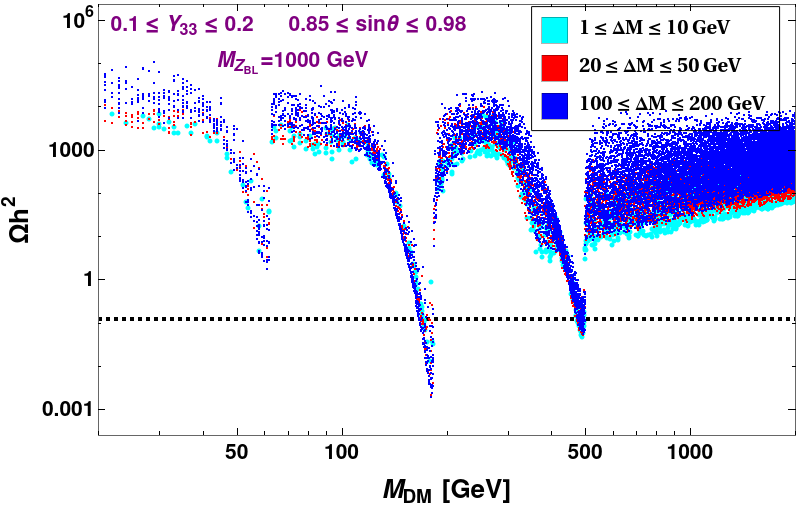}
			$$
			$$
			\includegraphics[scale=0.26]{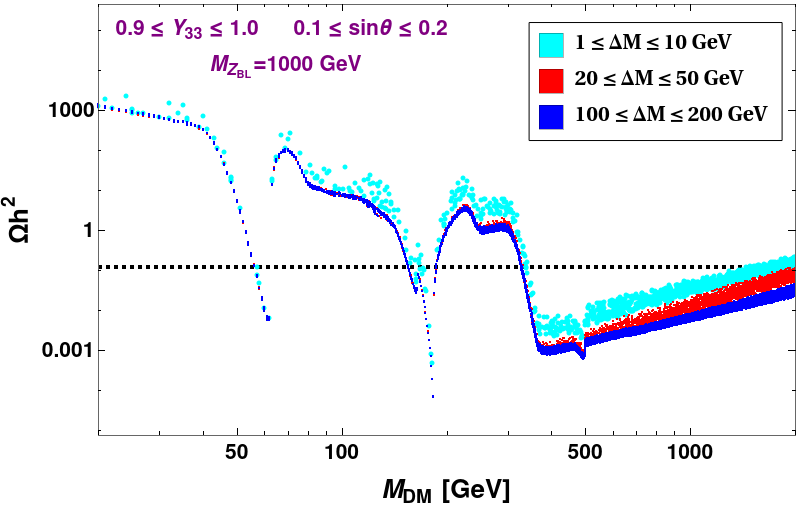}~
			\includegraphics[scale=0.26]{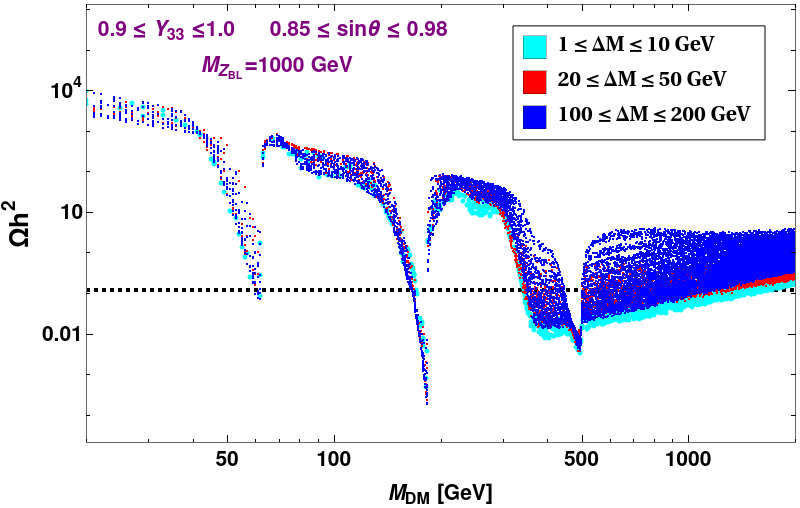}
			$$	
			\caption{Variation of relic density as a function of DM mass. Other parameters are kept fixed as
				mentioned inset of each figure.}
			\label{relic_M3}
		\end{figure}
		We show the variation of relic density of DM $\chi_{_3}$ in Fig.~\ref{relic_M3} 
		as a function of its mass $M_{DM}$ for different choices of $\Delta M$: 1-10 GeV, 20-50 GeV and 100-200 GeV shown by different colored points as mentioned in the inset of the figure. The dips in the relic density plots are essentially due to resonances corresponding to SM-Higgs, second Higgs and $Z_{\rm BL}$ gauge bosons respectively. In the top-left and top-right panel, $Y_{33}$ is varied in a range $0.1 \leq Y_{33} \leq 0.2$ whereas in the bottom-left and bottom-right panel it is varied in an interval $0.9 \leq Y_{33} \leq 1$. Clearly as $Y_{33}$ increases, the effective annihilation cross-section increases which decreases the relic density.
		
		
		\begin{figure}[h!]
			$$
			\includegraphics[scale=0.4]{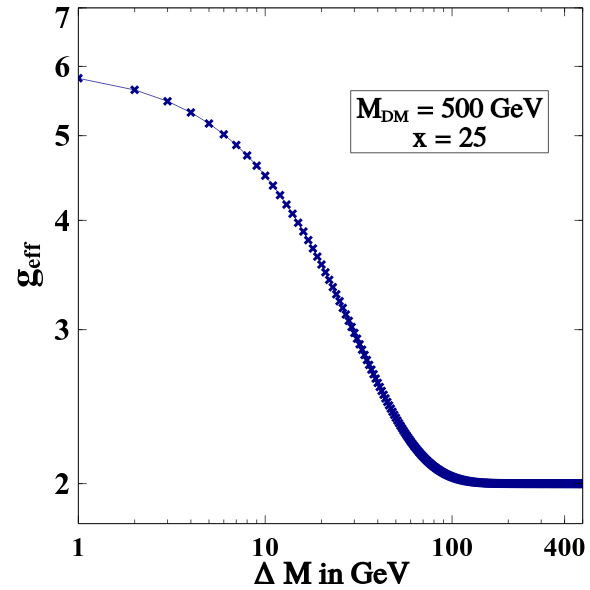}~~
			\includegraphics[scale=0.4]{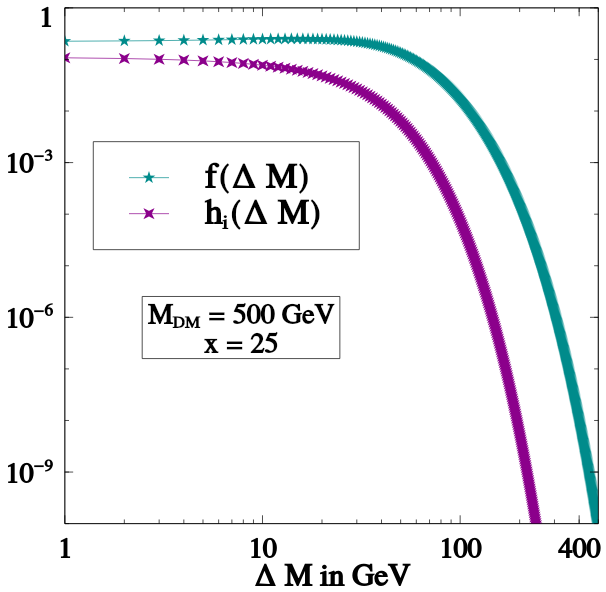}
			$$
			\caption{[Left]: Variation of $g_{eff}$ with $\Delta M$, [Right]: Variation of $f(\Delta M)=\frac{2g_3 g_1}{g^2_{eff}}\Big(1+\frac{\Delta M}{M_{DM}}\Big)^{\frac{3}{2}}\exp(-x \frac{\Delta M}{M_{DM}})$ and $h_{1,2}(\Delta M)=\frac{g^2_{1,2}}{g^2_{eff}}\Big(1+\frac{\Delta M}{M_{DM}}\Big)^{3}\exp(-2x \frac{\Delta M}{M_{DM}}) $ with $\Delta M$. }
			\label{geff_delm}
		\end{figure}
		
		We can also analyze the effect of mixing angle $\sin \theta$ and mass splitting ($\Delta M$) from the results in Fig~\ref{relic_M3}. As already mentioned, the parameter which decides the contribution of co-annihilations of DM to the relic density is $\sin\theta$ which can be understood by looking at Eq.~\ref{interaction_1} and \ref{interaction_2}. If $\sin\theta$ is small then the contribution from annihilation of DM will dominate overall co-annihilation effects but for larger $\sin \theta$, co-annihilation contributions will be more as compared to the annihilations. The value of $\sin \theta$ predominantly decides the relative contribution of annihilation and co-annihilations of DM for the calculation of relic density. However the mass-splitting also plays a crucial role in the effect of annihilations and co-annihilations of DM. With increase in mass-splitting, the contribution from co-annihilation processes gradually debilitates and becomes less effective. This is evident from Table.~\ref{st1tab},\ref{st7tab} and \ref{st9tab}.
		
		In the top and bottom right panel of Fig~\ref{relic_M3}, $\sin\theta$ is randomly varied in a range $0.85 \leq \sin\theta \leq 0.98$ for two different ranges of $Y_{33}$. For such a large $\sin\theta$, DM annihilates very weakly, so the co-annihilations essentially decides the effective annihilation cross-section and hence the relic density. This means in Eq.~\ref{sigmaveff}, the first term is negligible as compared to the other terms. In such a case, as $\Delta M$ increases, these co-annihilations become less and less effective thus decreasing the effective annihilation cross-section hence increasing the relic density. This trend is clearly observed in the right panel plots of Fig~\ref{relic_M3}. The effect of mass splitting in such a case can also be understood by looking at the right panel of Fig~\ref{geff_delm} where the multiplying functions (mentioned as $f(\Delta M)$ and $h(\Delta M)$) in the co-annihilation terms of the effective annihilation cross-section in Eq.~\ref{sigmaveff2}, are plotted as a function of mass-splitting $\Delta M$. As $\Delta M$ increases, these factors decreases drastically consequently decreasing the overall effective annihilation cross-section and hence increasing the relic density of DM.
		
		However if we consider the case of smaller $\sin\theta$ as considered for the left panel plots of Fig~\ref{relic_M3} ({\it i.e. $0.1 \leq \sin\theta \leq 0.2$}), here DM annihilation is the most effective and hence dominantly decides the relic density and except the first term in Eq.~\ref{sigmaveff}, other terms are negligible. In this case, with increase in mass splitting, the effective thermal averaged cross-section increases and relic density decreases. This is due to  the fact that, when $\Delta M$ increases, the effective degrees of freedom $g_{eff}$ decreases, which is shown in the left panel plot of Fig~\ref{geff_delm} for a benchmark value of $M_{DM}$. This, in turn, increases the $\langle \sigma v \rangle_{eff}$ and hence results in decrease in the DM relic abundance.
		
		The dominant number changing processes that can lead to the correct relic density for different DM mass is presented in 
		Table.~\ref{st1tab},\ref{st7tab} and \ref{st9tab} for three different values of $\sin\theta$ {\it i.e.} $\sin\theta=0.1,0.7 ~\&~ 0.9$ with dark 
		fermion mass splitting $\Delta M=1,10~\&~100$ GeV and $M_{Z_{\rm BL}}=1000$ GeV. If DM mass is smaller than $M_{W}$, it dominantly annihilates to SM 
		fermions through both Higgs and $Z_{\rm BL}$ exchange. As soon as kinematically allowed, DM then annihilates to $W^+W^-$ and $ZZ$ dominantly. Gradually 
		with the increase in DM mass, other channels involving additional scalars also open up. Once the DM mass is beyond the $Z_{\rm BL}$ threshold, it then dominantly annihilates to $Z_{\rm BL}$ and $H_3$. We can see from the Table.~\ref{st1tab}, that for $\sin\theta=0.1$, the dominant number changing processes are mostly the annihilation processes irrespective of the DM mass and mass splitting, however for $\sin\theta=0.9$ (Table.~\ref{st9tab}), it is mostly the co-annihilation processes that dominantly determine the relic. It is also worth noticing from these tables that the co-annihilations are most effective when $\Delta M$ is smaller and with increase in $\Delta M$, this effect gradually decreases.   
		
		\begin{table}[htb!]
			\renewcommand{\arraystretch}{1.4}
			\centering
			\resizebox{\linewidth}{!}{
				\begin{tabular}{|c|p{4.5cm}|p{4.5cm}|p{4.5cm}|}
					\hline \hline
					$M_{DM}$	& \multicolumn{3}{|c|}{Dominant Number Changing Processes }\\ 
					\cline{2-4}
					in GeV &~~~~$\Delta M=1$~~~~  &~~~~$\Delta M=10$~~~~&~~~~$\Delta M=100$~~~~ \\
					\hline
					30 & $\chi_{3}\chi_{3}\to f \bar{f}$(76\%)\newline$\chi_{1,2}\chi_{1,2}\to f \bar{f}$(12\%) & $\chi_{3}\chi_{3}\to f \bar{f}$ (100\%)&$\chi_{3}\chi_{3}\to f \bar{f}$(100\%)\\
					\hline
					100 &$\chi_{3}\chi_{3} \to W^+ W^-(51\%)\newline \chi_{3}\chi_{3} \to ZZ(19\%)\newline \chi_{3}\chi_{3} \to f\bar{f}(13\%)$ &$\chi_{3}\chi_{3} \to W^+ W^-(58\%)\newline \chi_{3}\chi_{3} \to ZZ(22\%)\newline \chi_{3}\chi_{3} \to f\bar{f}(20\%)$ &$\chi_{3}\chi_{3} \to W^+ W^-(58\%)\newline \chi_{3}\chi_{3} \to ZZ(22\%)\newline \chi_{3}\chi_{3} \to f\bar{f}(20\%)$\\
					\hline
					300 &$\chi_{3}\chi_{3}\to H_{1,3}H_3(19\%)\newline \chi_{3}\chi_{3}\to f \bar{f}(39\%)$\newline $\chi_2\chi_2\to f\bar{f}(20\%)\newline \chi_{1}\chi_{1}\to f\bar{f}(12\%)$  &$\chi_{3}\chi_{3}\to H_{1,3}H_3(28\%)\newline \chi_{3}\chi_{3}\to f \bar{f}(60\%)$ &$\chi_{3}\chi_{3}\to H_{1,3}H_3(31\%)\newline \chi_{3}\chi_{3}\to f \bar{f}(64\%)$\\
					\hline
					1000 &$\chi_{3}\chi_{3}\to Z_{\rm BL}H_3$(49\%)\newline $\chi_{2}\chi_{2}\to Z_{\rm BL}H_3$(23\%)\newline$\chi_{1}\chi_{1}\to Z_{\rm BL}H_3$(22\%) &$\chi_{3}\chi_{3}\to Z_{\rm BL}H_3$(75\%)\newline $\chi_{2}\chi_{2}\to Z_{\rm BL}H_3$(10\%)\newline$\chi_{1}\chi_{1}\to Z_{\rm BL}H_3$(9\%) &$\chi_{3}\chi_{3}\to Z_{\rm BL}H_3$(94\%)\newline $\chi_{3}\chi_{3}\to H_3H_3$(5\%)\\
					\hline
					\hline
				\end{tabular}
			}
			\caption{{Dominant annihilation and co-annihilation channels for $\pmb{\large \sin\theta=0.1}$ }}
			\label{st1tab}
		\end{table}
		\begin{table}[htb!]
			\renewcommand{\arraystretch}{1.4}
			\centering
			\resizebox{\linewidth}{!}{
				\begin{tabular}{|c|p{4.0 cm}|p{4.cm}|p{4.cm}|}
					\hline \hline
					$M_{DM}$	& \multicolumn{3}{|c|}{Dominant Number Changing Processes }\\ 
					\cline{2-4}
					in GeV &~~~~$\Delta M=1$~~~~  &~~~~$\Delta M=10$~~~~&~~~~$\Delta M=100$~~~~ \\
					\hline
					30 & $\chi_{1}\chi_{3}\to f \bar{f}$(71\%)\newline$\chi_{2}\chi_{2}\to f \bar{f}$(21\%) & $\chi_{3}\chi_{3}\to f \bar{f}$(57\%)\newline$\chi_{1}\chi_{3}\to f \bar{f}$ (30\%)\newline $\chi_{2}\chi_{2}\to f \bar{f}$ (6\%)&$\chi_{3}\chi_{3}\to f \bar{f}$(100\%)\\
					\hline
					100 &$\chi_{1}\chi_{3} \to f \bar{f}(23\%)$,\newline $\chi_{3}\chi_{3} \to W^+ W^-(17\%)$\newline $\chi_{1}\chi_{3} \to W^+ W^-(22\%)$,\newline $\chi_{3}\chi_{3} \to ZZ(6\%)\newline \chi_{1}\chi_{3} \to ZZ(9\%)$, \newline $\chi_{1}\chi_{1} \to W^+ W^-(8\%)$ &$\chi_{3}\chi_{3} \to W^+ W^-(54\%)\newline \chi_{3}\chi_{3} \to ZZ(20\%)\newline \chi_{1}\chi_{3} \to W^+ W^-(9\%)$ &$\chi_{3}\chi_{3} \to W^+ W^-(71\%)\newline \chi_{3}\chi_{3} \to ZZ(27\%)$\\
					\hline
					300 &$\chi_{1}\chi_{3}\to f \bar{f}(57\%), \newline \chi_{1}\chi_{3}\to H_{1,3}H_3(9\%)$\newline $\chi_2\chi_2\to f\bar{f}(21\%),\newline \chi_{3}\chi_{3}\to H_{1,3}H_3(5\%)\newline \chi_{1}\chi_{1}\to H_{1,3}H_3(5\%)$  &$\chi_{1}\chi_{3}\to f\bar{f}(54\%)\newline \chi_{2}\chi_{2}\to f \bar{f}(22\%)\newline \chi_{3}\chi_{3}\to H_{1,3}H_3(12\%)\newline \chi_{1}\chi_{3}\to H_{1,3}H_3(10\%)$ &$\chi_{3}\chi_{3}\to H_{1,3}H_3(64\%)\newline \chi_{1}\chi_{3}\to H_3H_3(20\%)\newline \chi_{3}\chi_{3}\to f \bar{f}(5\%)$\\
					\hline
					1000 &$\chi_{1}\chi_{3}\to Z_{\rm BL}H_3$(68\%)\newline $\chi_{2}\chi_{2}\to Z_{\rm BL}H_3$(28\%) &$\chi_{1}\chi_{3}\to Z_{\rm BL}H_3$(67\%)\newline $\chi_{2}\chi_{2}\to Z_{\rm BL}H_3$(27\%) &$\chi_{1}\chi_{3}\to Z_{\rm BL}H_3$(36\%)\newline $\chi_{3}\chi_{3}\to Z_{\rm BL}H_3$(31\%)\newline $\chi_{2}\chi_{2}\to Z_{\rm BL}H_3$(14\%)\newline $\chi_{3}\chi_{3}\to H_3H_3$(16\%)\\
					\hline
					\hline
				\end{tabular}
			}
			\caption{{Dominant annihilation and co-annihilation channels for $\pmb{\large \sin\theta=0.7}$ }}
			\label{st7tab}
		\end{table}

		\begin{table}[htb!]
			\renewcommand{\arraystretch}{1.4}
			\centering
			\resizebox{\linewidth}{!}{
				\begin{tabular}{|c|p{4cm}|p{4.cm}|p{4.cm}|}
					\hline \hline
					$M_{DM}$	& \multicolumn{3}{|c|}{Dominant Number Changing Processes }\\ 
					\cline{2-4}
					in GeV &~~~~$\Delta M=1$~~~~  &~~~~$\Delta M=10$~~~~&~~~~$\Delta M=100$~~~~ \\
					\hline
					30 & $\chi_{1}\chi_{3}\to f \bar{f}$(37\%),\newline $\chi_{2}\chi_{2}\to f \bar{f}$(37\%)\newline$\chi_{3}\chi_{3}\to f \bar{f}$(13\%) & $\chi_{3}\chi_{3}\to f \bar{f}$(74\%)\newline $\chi_{2}\chi_{2}\to f \bar{f}$ (23\%)&$\chi_{3}\chi_{3}\to f \bar{f}$(98\%)\\
					\hline
					100 &$\chi_{1}\chi_{3} \to f \bar{f}(12\%)\newline\chi_{1}\chi_{1} \to W^+ W^-(25\%)$\newline $\chi_{1}\chi_{3} \to W^+ W^-(15\%)\newline\chi_{1}\chi_{3} \to ZZ(6\%)\newline \chi_{2}\chi_{2} \to f\bar{f}(9\%)$\newline$\chi_{3}\chi_{3} \to W^+ W^-(3\%)$ &$\chi_{2}\chi_{2} \to f\bar{f}(23\%)\newline\chi_{3}\chi_{3} \to f\bar{f}(23\%)\newline \chi_{1}\chi_{1} \to f\bar{f}(6\%)\newline\chi_{3}\chi_{3} \to W^+ W^-(13\%)\newline \chi_{3}\chi_{3} \to ZZ(5\%)\newline \chi_{1}\chi_{3} \to W^+ W^-(9\%)$ &$\chi_{3}\chi_{3} \to f\bar{f}(58\%)\newline\chi_{3}\chi_{3} \to W^+ W^-(30\%)\newline \chi_{3}\chi_{3} \to ZZ(11\%)$\\
					\hline
					300 &$\chi_{1}\chi_{3}\to f \bar{f}(36\%)\newline\chi_{1}\chi_{3}\to H_{1,3}H_3(6\%)$\newline $\chi_2\chi_2\to f\bar{f}(24\%), \newline \chi_{1}\chi_{1}\to f\bar{f}(10\%)\newline \chi_{1}\chi_{1}\to H_{1,3}H_3(11\%)$  &$\chi_{1}\chi_{3}\to f\bar{f}(26\%)\newline \chi_{2}\chi_{2}\to f \bar{f}(36\%)\newline \chi_{3}\chi_{3}\to f \bar{f}(14\%)\newline  \chi_{1}\chi_{3}\to H_{1,3}H_3(10\%)$ &$\chi_{3}\chi_{3}\to H_{1,3}H_3(6\%)\newline \chi_{1}\chi_{3}\to H_3H_3(5\%)\newline \chi_{3}\chi_{3}\to f \bar{f}(78\%)$\\
					\hline
					1000 &$\chi_{1}\chi_{3}\to Z_{\rm BL}H_3$(41\%)\newline $\chi_{2}\chi_{2}\to Z_{\rm BL}H_3$(29\%)\newline $\chi_{3}\chi_{3}\to Z_{\rm BL}H_3$(11\%)\newline $\chi_{1}\chi_{1}\to Z_{\rm BL}H_3$(16\%) &$\chi_{1}\chi_{3}\to Z_{\rm BL}H_3$(34\%)\newline $\chi_{2}\chi_{2}\to Z_{\rm BL}H_3$(35\%) $\chi_{3}\chi_{3}\to Z_{\rm BL}H_3$(20\%)\newline $\chi_{1}\chi_{1}\to Z_{\rm BL}H_3$(9\%)&$\chi_{1}\chi_{3}\to Z_{\rm BL}H_3$(4\%)\newline $\chi_{3}\chi_{3}\to Z_{\rm BL}H_3$(70\%)\newline $\chi_{2}\chi_{2}\to Z_{\rm BL}H_3$(25\%)\\
					\hline
					\hline
				\end{tabular}
			}
			\caption{{Dominant annihilation and co-annihilation channels for $\pmb{\large \sin\theta=0.9}$ }}
			\label{st9tab}
		\end{table}

		\begin{figure}[htb!]
			$$
			\includegraphics[scale=0.27]{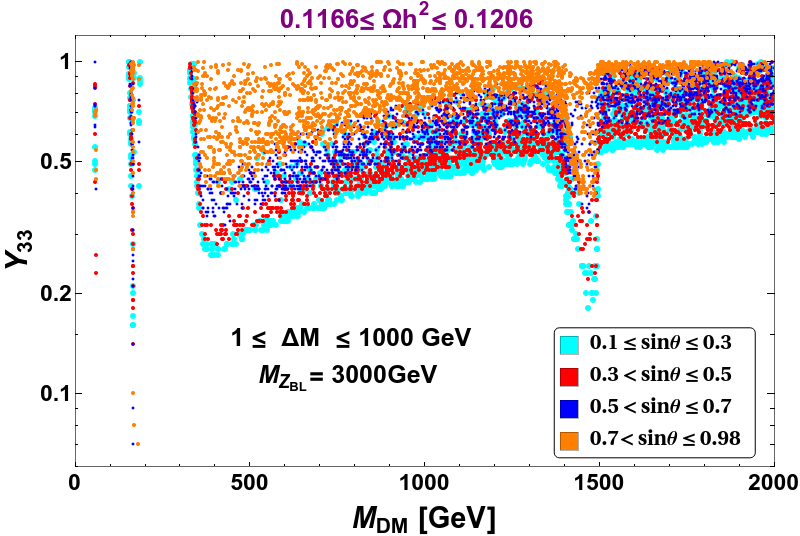}
			\includegraphics[scale=0.27]{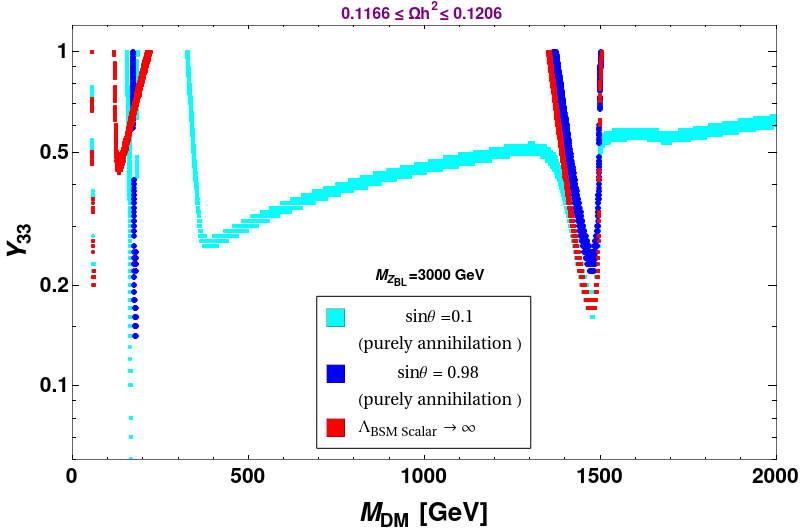}
			$$
			\caption{{[Left]:Relic density allowed parameter space in $M_{DM}-Y_{33}$ plane for different intervals of $\sin\theta$ where both annihilations and co-annihilations of DM are incorporated. [Right]:Relic density allowed parameters space in $M_{DM}-Y_{33}$ plane for different $\sin\theta$ values, where only annihilation of DM is considered in different limits.}
			}
			\label{y33_M3}
		\end{figure}

		To make the analysis more robust, in the left panel of Fig.~\ref{y33_M3}, the correct relic density allowed parameter space has been shown in the plane of 
		$Y_{33}$ vs $M_{DM}$ for wide range of mixing angle $\{\sin\theta= 0.1-0.3, 0.3-0.5, 0.5-0.7, 0.7-0.98\}$, indicated by different colors. To carry out 
		this scan of parameter space, $\Delta M$ is varied randomly within $1$ to $1000$ GeV.

		To establish the evidence of co-annihilations in generating the correct relic density in this scenario, one has to compare the left 
		and right panels of Fig. \ref{y33_M3}. In the right-panel of Fig~\ref{y33_M3}, we show the parameter space satisfying relic density constraint in the 
		plane of $Y_{33}$ vs $M_{DM}$, considering only the annihilation processes of the DM for three limiting cases: (i) small $\sin \theta$ limit, (ii) Higgs 
		decoupling limit and (iii) large $\sin \theta $ limit. \\

		$\bullet$ Case-I: Small $\sin \theta$ limit ($\sin \theta \to 0$)\\
		In this case the correct relic density is obtained by setting $\sin \theta =0.1$ as shown by the Cyan coloured points in the right panel of Fig. \ref{y33_M3}. 
		We see that apart from the resonances, the DM relic density can be satisfied for a wide range of DM mass with varying $Y_{33}$. This is essentially due to 
		the presence of additional Higgses and $Z_{\rm BL}$ in the theory. The annihilation of $\chi_3 \chi_3 \to H_i H_j,  H_i Z_{\rm BL}$ 
		can give rise to correct relic density beyond the resonances. As the DM mass increases, the relic density decreases which can be brought to the correct 
		ballpark by increasing the Yukawa coupling $Y_{33}$. This is exactly depicted by the cyan coloured points in Fig. \ref{y33_M3}.

		$\bullet$	Case-II: Higgs decoupling limit ($m_{H_i}\to \infty$)\\
		In this case the correct relic density is obtained by setting the masses of additional Higgses to a high scale. This is shown by the Red coloured points in the 
		right panel of Fig. \ref{y33_M3}. Except the Higgs masses, all other parameters are kept same as in case-I. In this case the dominant channels are $\chi_3 \chi_3 \to {\rm SM} {\rm SM}$ mediated by SM Higgs $H_1$ and $Z_{\rm BL}$. We see that the relic is satisfied only in the resonance regions. This clearly demonstrates that 
		in the small $\sin \theta$ limit(case-I) the additional Higgses only, allowing the DM mass beyond the resonance regime.

		$\bullet$ Case-III: Large $\sin \theta$ limit ($\sin \theta \to 1$)\\
		In this case the correct relic density is obtained by setting $\sin \theta =0.98$, while keeping all other parameters same as in case-I. We see from the 
		right panel of Fig. \ref{y33_M3} that the correct relic density is obtained only at the resonances as shown by the Blue points. This is because in the limit: 
		$\sin \theta \to 1$, the effective Yukawa coupling for annihilation goes to zero as shown in the left panel of Fig. \ref{coupling}. As a result the annihilation 
		cross-sections mentioned in case-I, i.e., $\chi_3 \chi_3 \to H_i H_j, H^+H^-, H^{++}H^{--}, H_i Z_{\rm BL}$ become small, leading to an overabundance of 
		$\chi_3$ outside the resonances. On the other hand, near resonances the cross-section increases, even if the annihilation coupling is small, and hence we get 
		correct relic of $\chi_3$.

		Remember that in the limit $\sin \theta \to 1$, the co-annihilation dominates over annihilation. See for instance, in the small $\Delta M$ limit 
		the processes: $\chi_i \chi_j \to \bar{f}f, W^+ W^-,H_i H_j, H_3 Z_{\rm BL}$ as given table \ref{st9tab}. Now if we incorporate all the number changing processes, 
		annihilations as well as co-annihilations, as done in the left panel Fig. \ref{y33_M3}, we see that the parameter space for correct relic density is 
		significantly enhanced. We get correct relic density of $\chi_3$ beyond the resonance regions. This confirms that for $\sin \theta \to 1$, 
		the co-annihilation dominates. Thus from this analysis, we can infer that the co-annihilations of the dark sector particles, do play a significant role in satisfying the correct relic density of the DM.

		\begin{figure}[htb!]
			$$
			\includegraphics[scale=0.4]{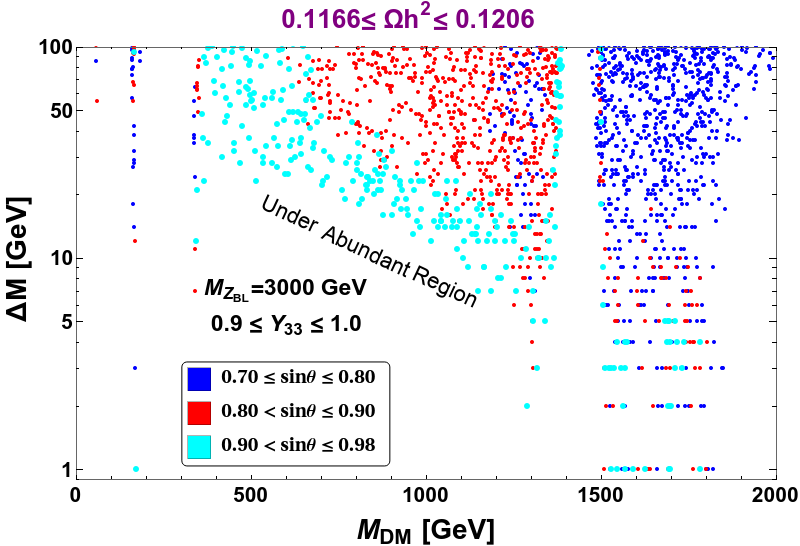}
			$$
			\caption{{Correct relic density satisfying points in the plane of $\Delta M$ and $M_{DM}$ for larger values of $\sin\theta$.}
			}
			\label{delm_M3}
		\end{figure}
		
		We also show the correct relic density satisfying points in the plane of $\Delta M$ and $M_{DM}$ in Fig.~\ref{delm_M3}, in the 
		large $\sin\theta$ range ($\sin\theta\in[0.7,~0.98]$), where the co-annihilation processes dominate over the annihilation processes. As previously discussed, the co-annihilation contributions are significant if the mass-splitting $\Delta M$ is not very large. For example, in the range $\Delta M = 1 - 20$ GeV and DM mass 
		in the range $1 - 1000$ GeV, the co-annihilation processes give rise a large cross-section on top of annihilation and thereby creating an under abundant 
		region. However, as $\Delta M$ increases, the co-annihlation cross-sections decreases. As a result, we get a correct relic density for DM mass 
		varying in the range $1 - 1000$ GeV. As we go from left to right, $\Delta M$ gradually decreases for a particular $\sin \theta$ to maintain the correct 
		relic density. For DM mass beyond 1000 GeV, the annihlation cross-sections decrease significantly. Therefore, we need a large co-annihilation 
		cross-section to give rise the relic density in right ballpark. This can be achieved by requiring a small $\Delta M$, typically $\Delta M < 10$ GeV. We 
		also see from Fig.~\ref{delm_M3} that, as we move from left to right while keeping $\Delta M$ fixed (preferably at $\Delta M > 50$ GeV), 
		large $\sin \theta$ favours a relatively small DM mass while small $\sin \theta$ prefers a large DM mass. This can be understood as follows. 
		When $\sin \theta$ is large, $\cos \theta$ is small, which indicate less annihilation. Therefore, we need 
		to increase the cross-section by choosing a relatively smaller DM mass to bring the relic density into the observed limit. On the other hand, when $\sin \theta$ is 
		small, $\cos \theta$ is large, which indicate larger annihilations and hence less relic. Therefore, the DM mass has to be increased in comparison to 
		large $\sin \theta$ limit to bring the relic density into the correct ballpark.   
		
		\section{Detection Prospects of DM}
		\label{DetCon}
		\subsection{Direct Detection}
		There are various attempts to detect DM. One such major experimental procedure is the Direct detection of the DM at  terrestrial laboratories through 
		elastic scattering of the DM off nuclei. Several experiments put strict bounds on the dark matter nucleon cross-section like LUX \cite{Akerib:2016vxi}, 
		PandaX-II \cite{Tan:2016zwf, Cui:2017nnn} and XENON-1T \cite{Aprile:2017iyp, Aprile:2018dbl}. In this model, the DM-nucleon 
		scattering is possible via Higgs-mediated interaction represented by the Feynman diagram shown in Fig~\ref{dmddfeyn}. Here, it is worth mentioning that 
		the DM being a Majorana fermion has only the off-diagonal (axial vector) couplings with the $Z_{\rm BL}$ boson and therefore do not contribute to spin independent direct search.
		
		\begin{figure}[h!]
			$$
			\includegraphics[scale=0.28]{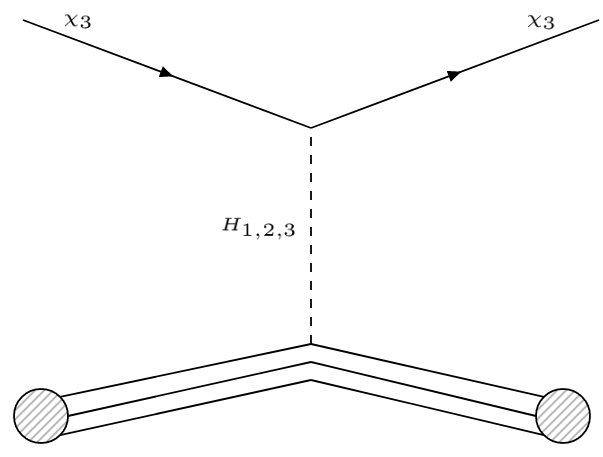}
			$$
			\caption{Higgs-mediated DM-nucleon scattering.}
			\label{dmddfeyn}
		\end{figure}
		
		The cross section per nucleon for the spin-independent (SI) DM-nucleon interaction is given by:
		\begin{equation}\label{dd1}
			\sigma_{SI} = \frac{1}{\pi A^2}\mu^2_r|\mathcal{M}|^2,
		\end{equation}
		where A is the mass number of the target nucleus, $\mu_r$ is the reduced mass of the DM-nucleon system and ${\mathcal M}$ 
		is the amplitude for the DM-nucleon interaction, which can be written as:
		\begin{equation}
			\label{dd2}
			\mathcal{M}=\Big[Z f_p +(A-Z)f_n\Big],
		\end{equation}
		where $f_{p}$ and $f_{n}$ denote effective interaction strengths of DM with proton and neutron of the target used with $A$ 
		being mass number and $Z$ is atomic number. The effective interaction strength can then further be decomposed in terms of interaction with partons as:
		\begin{equation}
			f_{p,n}^i = \sum_{q=u,d,s} f_{T_q}^{p,n} \alpha_q^{i} \frac{m_{p,n}}{m_q}+\frac{2}{27} f_{T_G}^{p,n}\sum_{Q=c,t,b} \alpha_Q^{i}\frac{m_{p,n}}{m_Q},
			\label{eq:coupling}
		\end{equation}  
		with 
		\begin{eqnarray}
			\alpha_q^{1} &=& -Y_{33}~\cos^2 \theta ~\frac{m_q}{v_H} \bigg[\frac{(s_{12} s_{23}-c_{12} c_{23} s_{13})^2}{m_{H_1}^2}\bigg]\nonumber\\
			\alpha_q^{2} &=& -Y_{33}~\cos^2 \theta ~\frac{m_q}{v_H} \bigg[\frac{( c_{12} s_{23}+c_{23} s_{12} s_{13})^2}{m_{H_2}^2}\bigg]\nonumber \\
			\alpha_q^{3} &=& -Y_{33}~\cos^2 \theta ~\frac{m_q}{v_H} \bigg[\frac{(c_{13}c_{23})^2}{m_{H_3}^2}\bigg]
			\label{eq: alpha}
		\end{eqnarray}
		coming from DM interaction with SM via Higgs portal coupling. In Eq.\ref{eq:coupling}, the different coupling strengths between DM and light quarks are given in ref. \cite{Bertone:2004pz,Alarcon:2012nr} as $f^p_{Tu} = 0.020 \pm 0.004, f^p_{Td} = 0.026 \pm 0.005, f^p_{Ts} = 0.014 \pm 0.062$, $f^n_{Tu} = 0.020 \pm 0.004, 
		f^n_{Td} = 0.036 \pm 0.005, f^n_{Ts} = 0.118 \pm 0.062$. The coupling of DM with the gluons in target nuclei is parameterized by:
		\begin{equation*}
			f^{(p,n)}_{TG} = 1- \sum_{q=u,d,s}f^{p,n}_{Tq}.
		\end{equation*}
		\begin{figure}[h!]
			$$
			\includegraphics[scale=0.4]{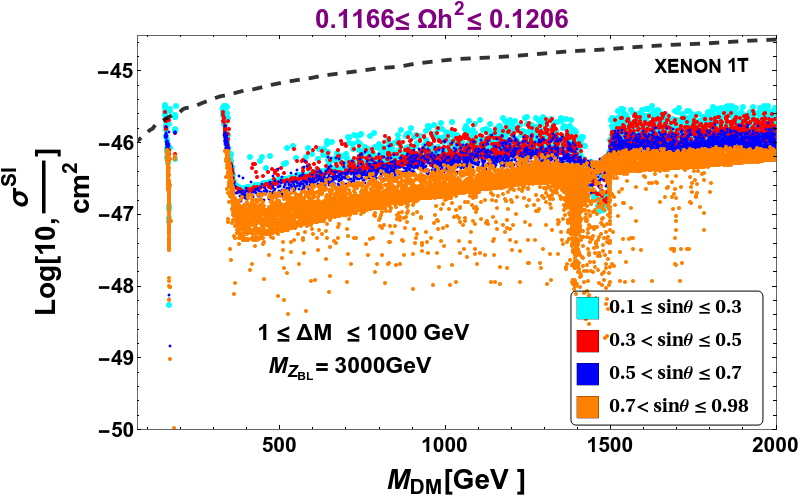}~
			$$
			\caption{Spin-independent direct detection cross section of DM with nucleon as a function of DM mass (in GeV) confronted with XENON-1T data over and above relic density constraint from PLANCK. }
			\label{DD_cr_M3}
		\end{figure}
		In the context of DM direct search, the model  parameters that enter the DM-nucleon direct search cross-section, are the Higgs-DM Yukawa coupling ($Y_{33}$) and the mixing angle ($\sin\theta$), which can be constrained by requiring that $\sigma_{SI}$ is less than the current DM-nucleon cross-sections dictated by non-observation of DM 
		in current direct search data. In Fig.~\ref{DD_cr_M3}, we show the DM-nucleon cross-section mediated by scalars in comparison to
		the latest XENON1T bound.
		In Fig.~\ref{DD_cr_M3}, we confronted the points satisfying relic density with the spin-independent DM-nucleon elastic scattering cross-section obtained for the model as a function of DM mass. The XENON1T bound is shown by dashed black line. Thus the region below this line satisfy both 
		relic density as well as direct detection constraint. We can see that, though for DM mass at the resonance regions, $\sin\theta$ values $0.1-0.98$ can satisfy the direct detection constraint but for DM masses other than at the resonances, only larger $\sin\theta$ values ($0.7\leq\sin\theta\leq0.98$) are favoured which is indicated by the orange points. As we have already discussed that in the larger $\sin\theta$ regime, the relic density is governed predominantly through the co-annihilations of DM, so this result interestingly implies that the co-annihilation effect essentially enhances the parameter space that satisfies the direct search constraints other than the resonance regions.


		\subsection{Indirect Detection}

		\begin{figure}[h!]
			$$
			\includegraphics[scale=0.4]{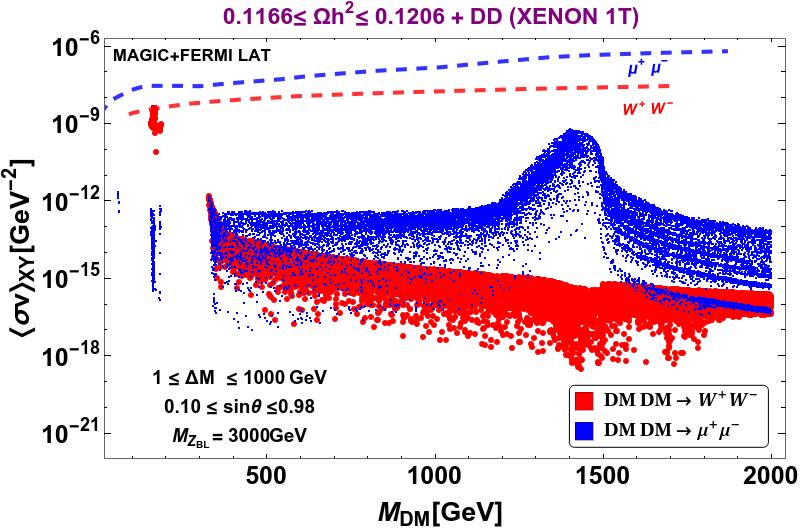}~
			$$	
			\caption{ {$ \langle \sigma v \rangle_{\chi_{3} \chi_{3} \rightarrow W^{+}W^{-}}$ and $ \langle \sigma v \rangle_{\chi_{3} \chi_{3} \rightarrow \mu^{+}\mu^{-}}$ are shown as a function of $M_{DM}$. Only the points satisfying both Relic and DD constraints are shown. Other parameters are kept fixed as mentioned in the inset of the figure.}}
			\label{ind_det_1}
		\end{figure}
		Apart from direct detection experiments, DM can also be probed at different indirect
		detection experiments which essentially search for SM particles produced through DM annihilations. Among these final states, photon and
		neutrinos, being neutral and stable can reach the indirect detection experiments without
		getting affected much by the intermediate medium between the source and the detector. For DM in the WIMP paradigm, these photons lie in the gamma ray regime and hence can be measured at space-based telescopes
		like the Fermi Large Area Telescope (FermiLAT) or ground-based telescopes like MAGIC. Measuring the gamma ray flux and using the standard astrophysical inputs, one can constrain the DM annihilation into different final states like $\mu^+ \mu^-, \tau^+ \tau^-, W^+ W^-, b\bar{b}$. Since DM can not couple to photons directly, gamma rays can be produced from such charged final states. Using the bounds on DM annihilation to these final states from the indirect detection bounds arising from the global
		analysis of the Fermi-LAT and MAGIC observations of dSphs~\cite{Ackermann:2015zua, Ahnen:2016qkx}, we check for the constraints on our DM parameters.
		
		Since there are multiple annihilation channels to different final
		states, the Fermi-LAT constraints on individual final states are weak for most of the cases. In Fig.~\ref{ind_det_1}, we show the points satisfying both relic constraint and direct search constraint confronted with the constraints from indirect detection from MAGIC+FermiLAT for annihilation of DM to $W^+~W^-$ and $\mu^+\mu^-$ which are the most stringent as compared to DM annihilation to other channels. In this model DM annihilation to $W^+~W^-$ can occur through scalar mediation as shown in right panel of Fig.~\ref{Feyn_diag4} and DM annihilation to $\mu^+~\mu^-$ can occur through scalar as well as gauge boson mediation as shown in  Fig.~\ref{Feyn_diag1}. The combined bound from MAGIC and FermiLAT are shown by the dashed lines. The points below these lines are allowed by relic, direct and indirect search constraints. 
		
		\subsection{Collider constraints on {$g_{\rm BL} - M_{Z_{\rm BL}}$}  }
		Apart from constraints from relic density and direct, indirect search of DM, there exists stringent experimental constraints on the $\rm B-L$ gauge sector from colliders like ATLAS,CMS and LEP-II. There exists a lower bound on the ratio of new gauge boson mass to the new gauge coupling $M_{Z'}/g' \geq 7$ TeV from LEP-II data~\cite{Carena:2004xs, Cacciapaglia:2006pk}. However the bounds from the current LHC experiments have already surpassed the LEP II limits. In particular, search for high mass
		Di-lepton resonances at ATLAS\cite{Aad:2019erb} and CMS\cite{Sirunyan:2021ctt} have put strict bounds on such additional gauge sector. In order to translate these constraints to our setup, we followed the strategy as mentioned in\cite{Das:2021esm} where the upper limit on the gauge coupling $g_{\rm BL}$ for a particular mass of gauge boson $M_{Z_{\rm BL}}$ can be derived as:
			\begin{equation}
				g^{\rm U.L.}_{\rm BL} = \sqrt{\frac{\sigma_{\rm Exp.}}{\sigma_{\rm Th.}/ g^2_{\rm Th.}}}
			\end{equation}  
			where $\sigma_{\rm Exp.}$ is the upper limit on the production cross-section of $p p \to Z_{\rm BL}+ X \to \ell^+ \ell^- +X~(\ell=e,\mu)$ and $\sigma_{\rm Th.}$ is the cross-section one obtains in their respective model for the same channel with corresponding gauge coupling $g^2_{\rm Th.}$.	We found that the constraint from ATLAS is more stringent than that from CMS, so we use the ATLAS ($\sqrt{s}=13~{\rm TeV}$ and $\mathcal{L}=139~fb^{-1}$) constraint to scrutinize the parameter space throughtout our analysis. Here it is worth mentioning that, because of the additional decay channels of $Z_{\rm BL}$ in our model as compared to the conventional $\rm B-L$ scenario, the derived constraints on $g_{\rm BL}-M_{Z_{\rm BL}}$ is relatively weaker as the branching fraction $Br(Z_{\rm BL} \to \ell^+ \ell^- )$ is relatively smaller.	

		\begin{figure}[htb!]
			$$
			\includegraphics[scale=0.4]{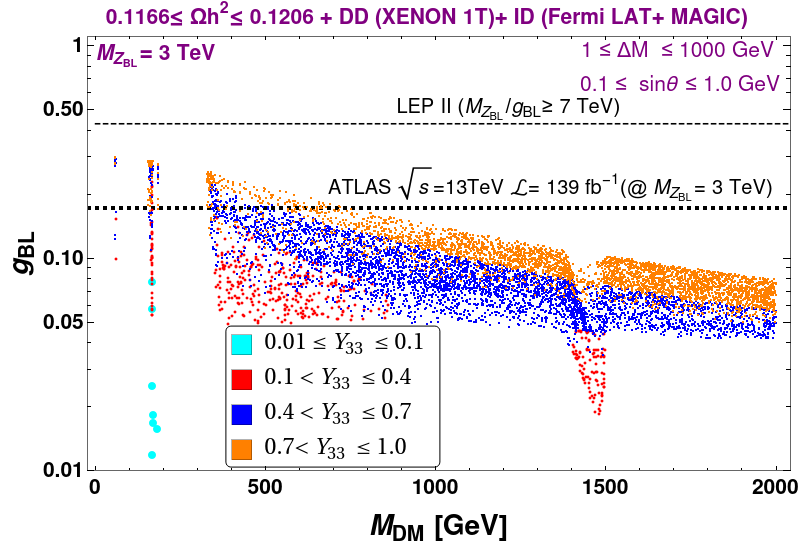}
			$$	
			\caption{{Parameter space satisfying Relic, Direct Detection and Indirect detection constraints are shown in $g_{\rm BL}$ vs $M_{DM}$ plot. ATLAS and LEP-II bounds are shown for $M_{Z_{BL}} = 3$ TeV by the black dotted lines}. }
			\label{gbl_m3}
		\end{figure}
		
		In Fig.~\ref{gbl_m3}, we show a parameter scan in the plane of $g_{\rm BL}$ vs $M_{DM}$ to scrutinize our parameter space with respect to the constraints from ATLAS and LEP-II. The bounds on $g_{\rm BL}$ for a fixed $M_{Z_{\rm BL}}$ from both LEP-II and ATLAS are shown by dotted black lines for $M_{Z_{\rm BL}}=3$ TeV. It is clear that the constraint from LEP-II is much weaker than the constraints from ATLAS. Only those points which lie below this black dotted line is allowed from all the relevant constraints({\it i.e.} Relic + Direct Detection + Indirect Detection + ATLAS). The different coloured points depict different $Y_{33}$ values. Here it is worth mentioning that for smaller values of $M_{Z_{\rm BL}}$ around 1 TeV, the constraint from ATLAS on the corresponding $g_{\rm BL}$ is more severe, thus ruling out most of the parameter space except at the resonances and regions beyond $M_{DM}$ $1$ TeV corresponding to $Y_{33}$ values larger than $0.4$. 
		However for larger values of $M_{Z_{\rm BL}}$, the corresponding constraint on $g_{\rm BL}$ from ATLAS, gradually debilitates and one can thus obtain more points satisfying all the relevant constraints.
		\vspace{1cm}
		~\\
		\underline{ \pmb{$2000$ GeV $\leq M_{Z_{\rm BL}} \leq 5000$ GeV}}:
		~\\
		
		
		\begin{figure}[htb!]
			$$
			\includegraphics[scale=0.4]{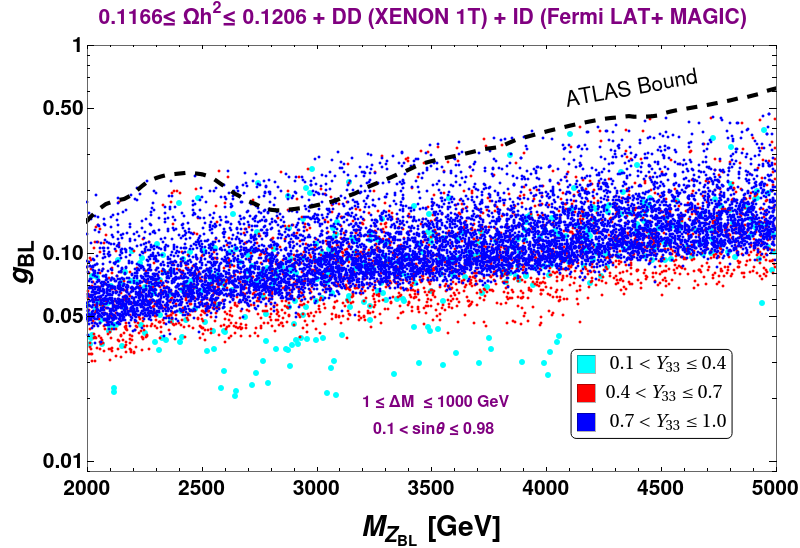}
			$$	
			\caption{
				Relic, direct detection and indirect detection satisfied points are shown in the plane of $g_{\rm BL}-M_{Z_{\rm BL}}$ plane  with different range of $Y_{33}$. The
				thick black dotted line shows the ATLAS bound on $g_{\rm BL} ~vs ~ M_{Z_{\rm BL}}$ plane from non-observation of $Z_{\rm BL}$ at colliders.
			}
			\label{zblvaried_1}
		\end{figure}
		So far whatever analysis we have done is with a fixed mass of the $\rm B-L$ gauge boson. We now turn to find the allowed parameter space in light of ATLAS
		bound on $g_{\rm BL}-M_{Z_{\rm BL}}$. The constraint on $g_{BL}$ for corresponding values of $M_{Z_{BL}}$ coming from the non-observation of a new gauge boson ($Z_{\rm BL}$) at LHC from ATLAS~\cite{Aaboud:2017buh} analysis 
		is shown by the black thick dotted line in Fig.~\ref{zblvaried_1}. This indicates that points below the line with smaller $g_{BL}$ is allowed, 
		while those above the line are ruled out. The plot shows points that satisfy relic density constraint, direct as well as indirect search constraints. Different colours indicate ranges of $Y_{33}$ as mentioned in figure inset.
		
		\begin{figure}[h!]
			$$
			\includegraphics[scale=0.28]{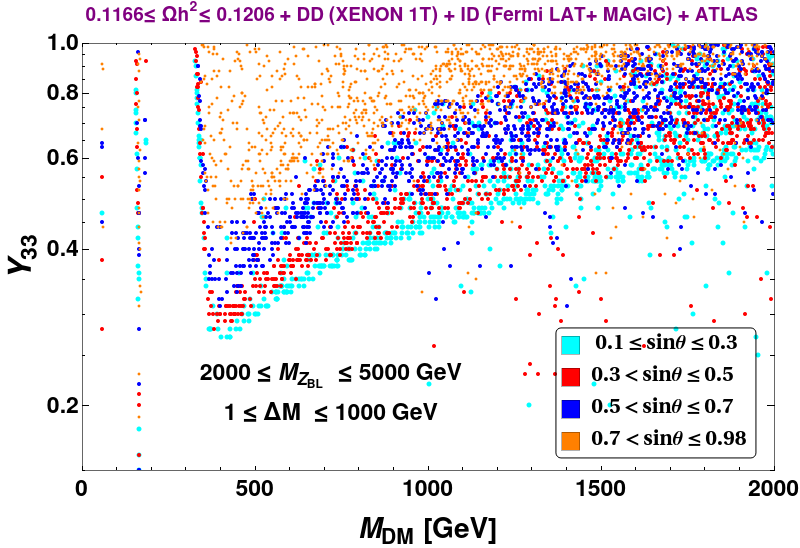}
			\includegraphics[scale=0.28]{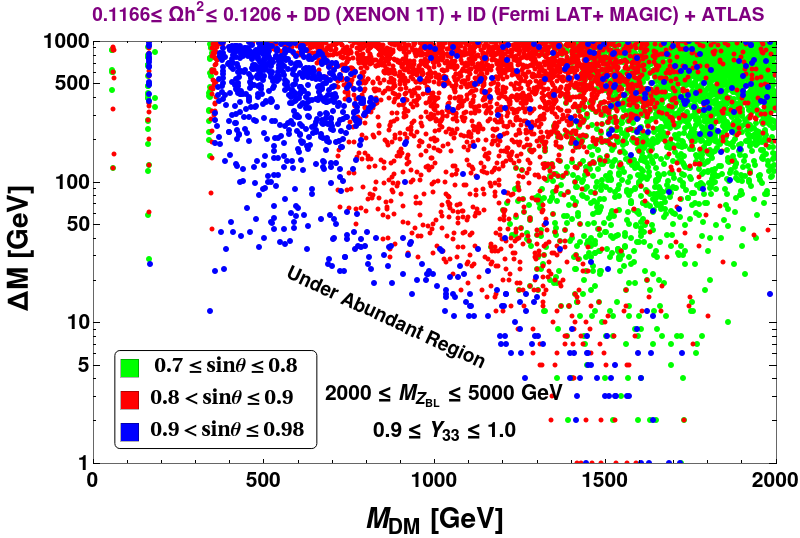}
			$$ 	
			\caption{{[Left]:Parameter space satisfying relic density, direct and indirect detection bound as well as $g_{\rm BL} - M_{Z_{\rm BL}}$ constraint from ATLAS is shown in the plane of $Y_{33}~vs~M_{DM}$ [Right]: Parameter space satisfying all constraints in the plane of $\Delta M$ and $M_{DM}$ for larger values of $\sin\theta$.}}
			\label{finalps}
		\end{figure}

		We then showcase the final parameter space in Fig.\ref{finalps}. In the left panel we represent the points in the plane of $Y_{33}$ vs $M_{DM}$ after imposing the bounds from correct relic density of DM, direct and indirect detection of DM and search for $\rm B-L$ gauge boson at ATLAS. Clearly there is enough parameter space beyond the resonance regions that is allowed from all the relevant constraints. Also the points with larger $\sin\theta$ which represents the dominant co-annihilation of dark sector fermions play a significant role in giving correct relic density as well as satisfying all other phenomenological and experimental constraints. 
		
		To specifically depict the parameter space where the co-annihilations do play a significant role, we show the parameter space with larger dark fermion mixing angle ($\sin\theta \in [0.7,1]$), in the plane of $\Delta M ~vs~ M_{DM}$. Clearly, for $\sin\theta \to 1$(Blue coloured points), as we increase the DM mass, $\langle \sigma v \rangle_{eff}$ decreases which can be compensated by the help of more co-annihilation contributions that can be achieved by decreasing mass-splittings.
		
		\section{Collider Signature of Doubly Charged Scalar in Presence of $Z_{BL}$}
		\label{collider}
		\begin{figure}[h!]
			$$
			\includegraphics[scale=0.4]{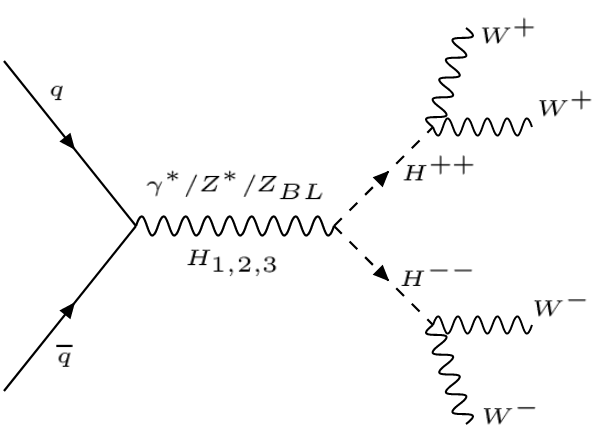}
			$$
			\caption{Feynman diagram for pair production of doubly charged scalar at LHC via Higgs ($H_{1,2,3}$) and Gauge Bosons ($\gamma, Z,Z_{\rm BL}$) mediations.  } 
			\label{feyncoll}
		\end{figure}
		
		The light doubly charged scalar in this model offers novel multi-lepton signatures with missing energy and jets. It is worthy of mentioning here that the dark sector which contains the gauge singlet Majorana fermions do not have any promising collider signatures as the mono-X type signal processes arising out of initial state radiation are extremely suppressed. The doubly charged scalar, $H^{\pm\pm}$ which is also charged under $U(1)_{\rm B-L}$ can be produced at Large Hadron Collider (LHC)  via Higgs ($H_{1,2,3}$) and gauge bosons ($\gamma, ~Z,~ Z_{BL}$)  mediations. Further decay of $H^{\pm\pm}$ to $W^\pm W^\pm$ pair ( assumed $m_{H^{\pm\pm}} \geq 2 m_W$) with almost $100\%$ branching ratio for $v_{\xi} \sim ~2.951$ GeV yields $W^+ W^+ W^-W^-$ final state. As a result the four $W$ final state offers: $4 \ell + \slashed{E}_T$ and $m~ \ell + n~ j + \slashed{E}_T$ signatures at collider. For details of branching fraction and partial decay widths of $H^{++}$ with $v_\xi$, please see appendix~\ref{hppdec}. Although this type of signatures have been studied in the context of Type -II seesaw model,  the triplet scalar $\xi$ considered in this model also have $U(1)_{\rm B-L}$ charge on top SM gauge charges and that makes this model different from the usual type-II seesaw scenario. In this section we will briefly highlight the effect of additional gauge boson $Z_{\rm BL}$ on the pair production cross-section of doubly charged scalar. The corresponding Feynman diagram of this type process is shown in Fig.\ref{feyncoll}.
		\begin{figure}[htb!]
			$$
			\includegraphics[scale=0.35]{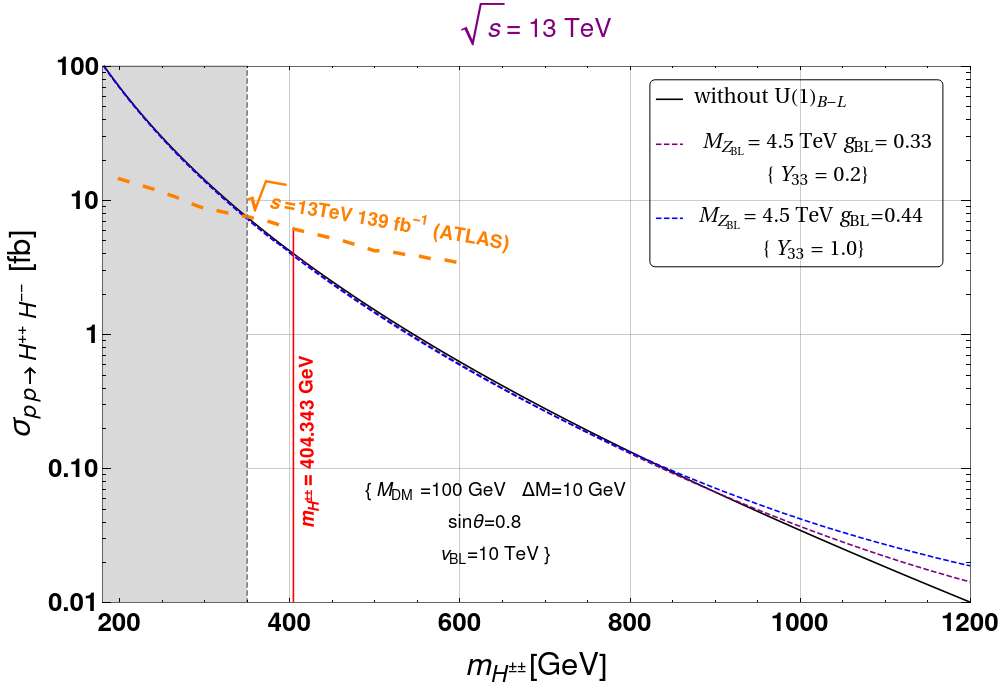}
			$$
			\caption{ Production cross-section for $p~p \to H^{++}~ H^{--}$ as a function of doubly charged scalar mass $m_{H^{\pm\pm}}$ considering Br($H^{\pm\pm} \to W^\pm W^\pm$) $\sim 100 \%$ at $\sqrt{s}=14$ TeV. The black solid line corresponds to the usual type-II seesaw scenario where $Z_{BL}$ gauge mediated diagrams are absent. The effects of $Z_{BL}$ on the production cross-section are shown dashed in dashed lines: purple line ($M_{Z_{\rm BL}}=4.5$ TeV, $g_{\rm BL}=0.33$) and blue line ($M_{Z_{\rm BL}}=4.5$ TeV, $g_{\rm BL}=0.44$). Other parameters are fixed as mentioned inset of the figure. The shaded region is excluded from ATLAS data on doubly charged scalar mass, $m_{H^{\pm\pm}}$ for $\sqrt{s}=13$ TeV and luminosity 139 fb$^{-1}$. } 
			\label{fig:feyncoll}
		\end{figure}
		
		The pair production cross-section of doubly charged scalar, $ H^{++} H^{--}$  as function of mass, $m_{H^{\pm\pm}}$ for fixed value of $M_{Z_{\rm BL}}=4.5$ TeV with $\sqrt{s}=13$ TeV is shown in Fig.\ref{fig:feyncoll}. The production cross-sections are computed in {\texttt{MicrOmegas}}  using the NNPDF23 parton distributions. The black solid line corresponds to the case where $U(1)_{\rm B-L}$ gauge boson, $Z_{\rm BL}$ is absent and the scenario resembles the usual type-II seesaw scenario. And in that case the $H^{++} H^{--}$ pair can be produced via SM Higgs  and SM gauge boson ($\gamma, Z$) mediated Drell-Yan processes. However in a gauged $\rm B-L$ scenario, the presence of the additional gauge boson $Z_{\rm BL}$ can affect this pair production cross-section of $H^{++} H^{--}$.  The effects of $U(1)_{\rm B-L}$ gauge boson on top of SM gauge bosons are shown by dotted lines in the Fig.\ref{fig:feyncoll} for two different values of gauge couplings: $g_{BL}=0.33$ (purple line) and $0.44$ (blue line) . It is important to note here that the above values of the $g_{BL}$ can be obtained using the Eqn.\ref{gbl} keeping the other parameters fixed as mentioned in the inset of the figure. For illustration purpose we considered two moderate values of $g_{\rm BL}$  : $0.33$ (purple dashed line) and $0.44$ (blue dashed line) which are in agreement with the current ATLAS bound $g_{\rm BL} \leq 0.47 $ for $M_{Z_{\rm BL}}=4.5$ TeV. It is noticeable from the graph that the presence of $Z_{\rm BL}$ enhances the production cross-section towards the heavy mass region of doubly charged scalar with moderate value of $g_{\rm BL}$ compared to the case without $U(1)_{\rm B-L}$ augmentation. It is because of the on-shell decay of $Z_{\rm BL}$ to $H^{++} H^{--}$ pair as $M_{Z_{\rm BL}} > 2 m_{H^{\pm\pm}}$ and there is a constructive interference between $Z_{\rm BL}$ and the SM gauge bosons. The orange dashed line shows the observed and expected upper limit on $H^{++}H^{--}$ pair production cross-section as a function of doubly charged scalar mass $m_{H^{\pm\pm}}$ at $95\%$ CL which is obtained from the combination of multi-leptons with jets plus missing energy search at ATLAS with $\sqrt s =13$TeV and integrated luminosity, $\mathcal{L}=139$ fb$^{-1}$~\cite{atlas}. This upper limit of production cross-section excludes the region of doubly charged triplet mass,$m_{H^{\pm\pm}} \leq 350$ GeV as shown by the shaded region in Fig.\ref{fig:feyncoll}.
		
		\begin{figure}[htb!]
			$$
			\includegraphics[scale=0.3]{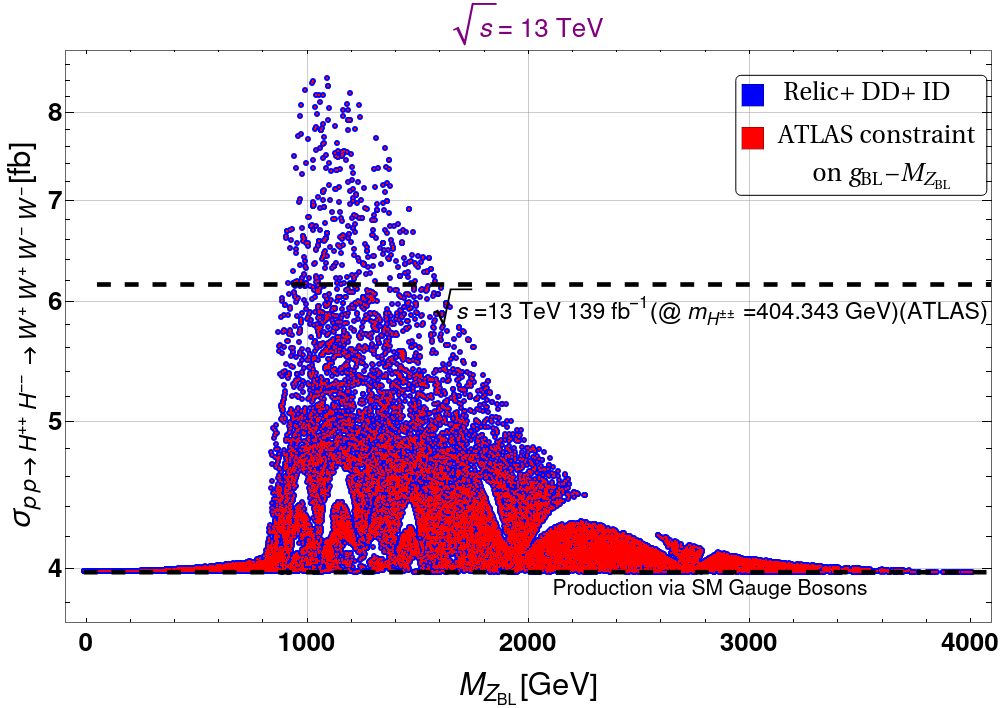}
			$$	
			\caption{The model parameters are projected against the production cross-section of doubly charged scalar ($H^{\pm\pm}$) as a function of $\rm B-L$ gauge boson mass $M_{Z_{\rm BL}}$. The blue points are satisfying the Relic+Direct Detection+ Indirect Detection constraints. Red points are consistent with the ATLAS constraint on $g_{\rm BL}-M_{Z_{\rm BL}}$.  }
			\label{prdcrszbl}
		\end{figure}
		
		In figure~\ref{prdcrszbl}, we projected the points satisfying all the relevant constraints against the doubly charged scalar ($H^{\pm\pm}$) production cross-section as a function of $\rm B-L$ gauge boson mass $M_{Z_{\rm BL}}$ with $\sqrt{s}=$ 13 TeV for a benchmark value of $m_{H^{\pm \pm}}=404.343$ GeV. The black dashed line shows the upper limit on the production cross-section from ATLAS~\cite{atlas}. The blue points show the parameters that satisfy all the relevant constraints like correct relic density, direct and indirect search of DM and the red points are obtained after imposing the constraints from ATLAS on $g_{\rm BL}$ and $M_{Z_{\rm BL}}$. We can see that in presence of the $\rm B-L$ gauge boson, the production cross-section $\sigma_{pp \to H^{++}H^{--}}$ can get a distinctive enhancement as compared to the case where production of $H^{\pm \pm}$ happens through SM gauge bosons ($\gamma^*, Z^*$) mediation only which is shown as the dashed black line at the bottom for easy comparison. As is clear from the Fig.~\ref{prdcrszbl}, near the resonance ({\it i.e.} $M_{Z_{\rm BL}} = 2 m_{H^{++}}$), we see maximum enhancement in the production cross-section which is again constrained from the $4 W$ final state at ATLAS and the points above the orange dotted line are ruled out.  
		
		
		\begin{figure}[htb!]
			$$
			\includegraphics[scale=0.3]{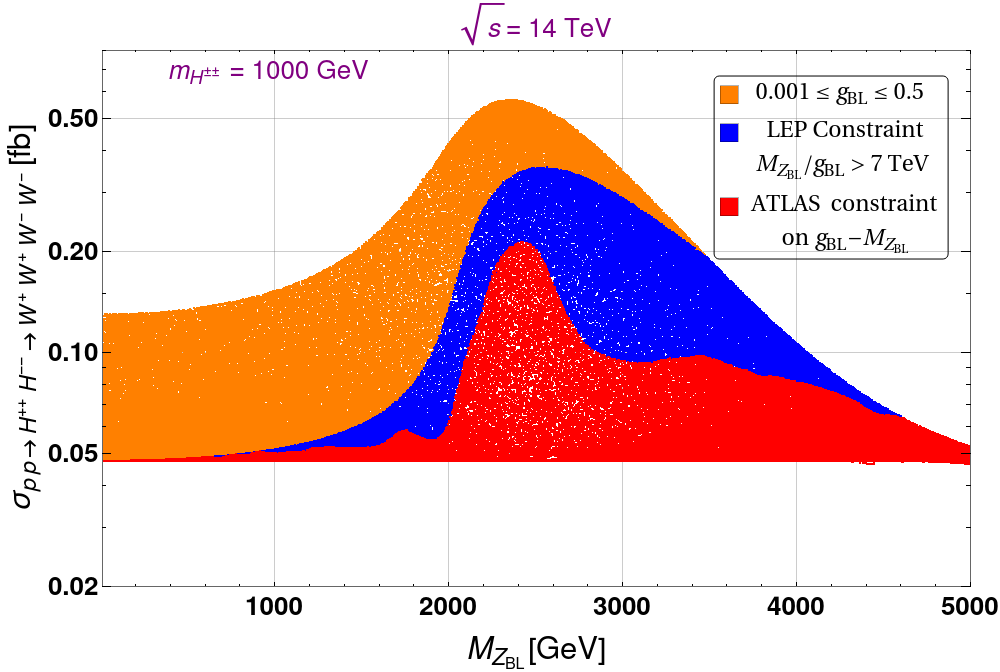}
			$$	
			\caption{ The production cross-section of doubly charged scalar ($H^{\pm\pm}$) as a function of  $M_{Z_{\rm BL}}$. The orange points correspond to gauge coupling:$0.001 \leq g_{\rm BL} \leq 0.50$. The blue points are allowed from LEP exclusion bound. Red points are consistent with ATLAS constraint on $g_{\rm BL}-M_{Z_{\rm BL}}$.  }
			\label{prdcrszbl1}
		\end{figure}
		
		Similar perceptible signal can be seen at the collider if we consider the doubly charged scalar mass in the TeV scale too, requiring a higher $M_{Z_{\rm BL}}$ ($> 2$ TeV) for the resonance enhancement to happen. Thus to demonstrate this fact, we considered the doubly charged scalar mass  $m_{H^{\pm\pm}}=1$ TeV. In figure~\ref{prdcrszbl1}, we show the  production cross-section of doubly charged scalar ($H^{\pm\pm}$) as a function of $M_{Z_{\rm BL}}$ considering the gauge coupling within the interval $0.001 \leq g_{\rm BL} \leq 0.50$ with $\sqrt{s}=$ 14 TeV shown by the orange points. Though the constraints from the current LHC experiments have already surpassed the LEP II limits on $g_{\rm BL}-M_{Z_{\rm BL}}$, for comparison we show the blue points in the plot which depicts the maximum increase in $\sigma_{pp \to H^{++}H^{--}}$ when the constraint from LEP on $g_{\rm BL}-M_{Z_{\rm BL}}$ is incorporated into the calculation. However, even after imposing the most stringent constraint from ATLAS on $g_{\rm BL}-M_{Z_{\rm BL}}$, we observe that there is a noteworthy enhancement in the production of $H^{\pm\pm}$ as compared to the value predicted by SM. The production cross-section $\sigma_{pp \to H^{++}H^{--}}$ increases by almost $300\%$ (0.21 fb in presence of $Z_{\rm BL}$ as compared to 0.047 fb predicted by SM) at the resonance. Apart from resonance also, there is significant enhancement for other masses of $Z_{\rm BL}$; for example, we see an enahncement by almost $90\%$ ($\sim 0.09$ fb in presence of $Z_{\rm BL}$ as compared to 0.047 fb predicted by SM) for $M_{Z_{\rm BL}}$ around $3.5$ TeV. This feature is evident from the red points in fig~\ref{prdcrszbl1}. This is the crucial evidence of the scenario considered here that can be probed by the near and future colliders and hence the feasibility of this model can be verified.   }
	
	This also establishes an interesting connection between the dark sector and the generation of neutrino mass via the modified type-II seesaw in a gauged $\rm B-L$ setup that we discussed.
	\section{Summary and Conclusions}
	\label{conclusion}
	
	In this paper, we have studied a very well motivated gauge extension of the standard model by augmenting the SM gauge group with a $U(1)_{\rm B-L}$ 
	symmetry, which happens to be an accidental symmetry of SM, to simultaneously address non-zero masses of light neutrinos as well as a 
	phenomenologically viable dark matter component of the universe. We minimally extend the fermion particle content of the model by adding 
	three exotic right chiral fermions $\chi_{i_R}$ with $\rm B-L$ charges $-4,-4$ and $+5$ in order to cancel the gauge and gravitational 
	anomalies that arise when one gauges the $\rm B-L$ symmetry. The stability of these fermions is owed to the remnant $\mathcal Z_2$ symmetry 
	after the $U(1)_{\rm B-L}$ breaking which distinguishes the added fermions from the SM as $\chi_{i_R} (i=e, \mu, \tau)$  are odd under $\mathcal Z_2$ 
	while all other particles are even. Thus the dark matter emerges as the lightest Majorana fermion from the mixture of these exotic fermions.
	
	A very interesting and important aspect of the model is the correlation between dark sector and neutrino mass generation. The neutrino mass is 
	explained through a modified type-II seesaw at TeV scale by introducing two $SU(2)_{L}$ triplet scalars $\Delta$ and $\xi$. $\Delta$ is super heavy 
	and doesn't have a coupling with the lepton and hence can not generate the neutrino mass even after acquiring an induced vev after the EW symmetry 
	breaking. Thus the neutrino mass is essentially generated through the $\xi L L$ coupling as given in Eq.~\ref{numasseqn}. In the limit $ v_{\rm BL} 
	\rightarrow 0$, which essentially means vanishing mixing between $\Delta$ and $\xi$, the neutrino mass also vanishes. Also Eqs.~\ref{lagrangian}, \ref{fermion_mass_matrix} and ~\ref{dm_mixing_angle} implies that the interactions between $\chi_{\tau_R}$ and $\chi_{e_R},\chi_{\mu_R}$ are established 
	through the scalar $\Phi_{\rm BL}$. In the limit of $\langle \Phi_{\rm BL} \rangle$ $\rightarrow 0$,  which essentially implies $\sin\theta\rightarrow 0$, 
	the DM candidate $\chi_3$ decouples from the heavier dark particles $\chi_1$ and $\chi_2$. In this limit there will be no co-annihilations among the dark 
	sector particles. Thus only if $\langle \Phi_{\rm BL} \rangle \neq 0$, we get non-zero masses of light neutrinos as well as it switches on the 
	co-annihilations of DM and hence enlarges the parameter space satisfying all relevant constraints. 
	
	We studied the model parameter space by taking into account all annihilation and co-annihilation channels for DM mass ranging from 1 GeV to 2 TeV. We confronted 
	our results with recent data from PLANCK and XENON-1T to obtain the parameter space satisfying relic density as well as direct detection constraints. The 
	DM being Majorana in nature, it escapes from the gauge boson mediated direct detection constraint. We also checked for the constraints on our model 
	parameters from the indirect search of DM using the recent data from Fermi-LAT and MAGIC which we found to be relatively weaker than other constraints. We 
	also imposed the constraint on $g_{\rm BL}- M_{Z_{\rm BL}}$ from current LHC data to obtain the final parameter space allowed from all constraints 
	including correct relic, direct and indirect detection of DM as well as the constraints from colliders on the $\rm B-L$ gauge boson and the corresponding 
	coupling.
	
	We also studied the detection prospects of the doubly charged scalar triplet which can have novel signatures at the colliders with multi-leptons along with missing energy and jets. We showed how in the presence of the $\rm B-L$ gauge boson $Z_{\rm BL}$, the pair production cross-section of $H^{++} H^{--}$ can get enhanced 
	and also depicted how the dark parameters satisfying all the relevant constraints can affect the production of this doubly charged scalar. 
	\section*{Acknowledgements}\label{ack}
	SM would like to acknowledge Alexander Belyaev and Alexander Pukhov for useful discussions. SM also thanks A. Das and P.S.B. Dev for useful discussions. PG would like to acknowledge the support from DAE, India for the Regional Centre for Accelerator based Particle Physics (RECAPP), Harish Chandra Research Institute. NS acknowledges the support from Department of
	Atomic Energy (DAE)- Board of Research in Nuclear
	Sciences (BRNS), Government of India (Ref. Number:
	58/14/15/2021- BRNS/37220).
	\appendix
	
	\section{Anomaly Cancellation}\label{anomaly}
	In any chiral gauge theory the anomaly coefficient is given by~\cite{palbook}:
	\begin{equation}
		\mathcal{A}=Tr[T_a[T_b,T_c]_+]_R-Tr[Ta[T_b,T_c]_+]_L\,,
	\end{equation} 
	where $T$ denotes the generators of the gauge group and $R$, $L$ represent the interactions of right and left chiral fermions with the gauge bosons.
	
	Gauging of $U(1)_{\rm B-L}$ symmetry within the SM lead to the following triangle anomalies: 
	\begin{equation*}
		\mathcal{A}_1[U(1)^3_{\rm B-L}]=3
	\end{equation*}
	\begin{equation}
		\mathcal{A}_{2}[(Gravity)^2 \times U(1)_{\rm B-L}]=3\,.
	\end{equation}
	The natural choice to make the  gauged $\rm B-L$ model anomaly free is by introducing three right handed neutrinos, each of having $\rm B-L$ charge $-1$ such that they result in $\mathcal{A}_1[U(1)^3_{\rm B-L}]=-3$ 
	and $\mathcal{A}_{2}[(Gravity)^2 \times U(1)_{\rm B-L}]=-3$ which lead to cancellation of above mentioned gauge anomalies. 
	
	However we can have alternative ways of constructing anomaly free versions of $U(1)_{\rm B-L}$ extension 
	of the SM. In particular, 
	three right chiral fermions with exotic $\rm B-L$ charges -4,-4,+5 can also give rise to vanishing $\rm B-L$ anomalies.
	\begin{equation*}
		\mathcal{A}_1[U(1)^3_{B-L}]=\mathcal{A}^{SM}_1[U(1)^3_{B-L}]+\mathcal{A}^{New}_1[U(1)^3_{B-L}]= 3+[(-4)^3+(-4)^3+(5)^3]=0
	\end{equation*}
	\begin{align}
		\mathcal{A}_{2}[(Gravity)^2 \times U(1)_{B-L}]&=\mathcal{A}^{SM}_{2}[(Gravity)^2 \times U(1)_{B-L}]
		+\mathcal{A}^{New}_{2}[(Gravity)^2 \times U(1)_{B-L}]\nonumber\\ 
		&=3+[(-4)+(-4)+(5)]=0
		\label{anomaly_cancel}
	\end{align} 
	
	\section{Decay of Doubly Charged Scalar}\label{hppdec}
	\begin{figure}[htb!]
		$$
		\includegraphics[scale=0.5]{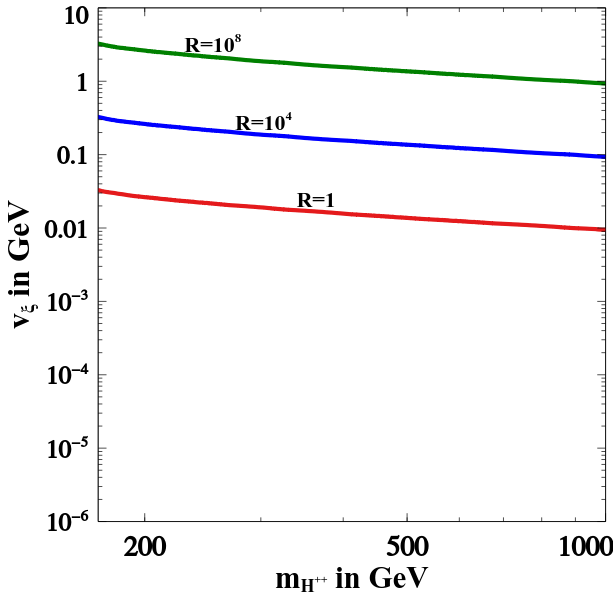}
		$$	
		\caption{\footnotesize{ The contour for R in the plane of $v_{\xi}~vs~m_{H^{++}}$. }}
		\label{Xi_decay_contour}
	\end{figure}
	The partial decay widths of the doubly charged scalar($H^{++}$) are given as:
	\begin{equation}
		\Gamma(H^{++} \rightarrow l^{+}_{\alpha}l^{+}_{\beta})= \frac{m_{H^{++}}}{4 \pi v^2_{\xi}(1+\delta_{\alpha \beta})} {\vert (M_{\nu})_{\alpha \beta}} \vert^2
	\end{equation}
	and 
	\begin{equation}
		\Gamma(H^{++}\rightarrow W^{+}W^{+})=g^4 v^2_{\xi} m^3_{H^{++}} \Big[1-4\Big(\frac{m_W}{m_{H^{++}}}\Big)^2\Big]^{\frac{1}{2}}\Big[1-4\Big(\frac{m_W}{m_{H^{++}}}\Big)^2+\Big(\frac{m_W}{m_{H^{++}}}\Big)^4\Big]
	\end{equation}

	This can be well analyzed by plotting contours of the ratio
	\begin{equation}
		R = \frac{\Gamma(H^{++}\rightarrow W^{+}W^{+})}{\Gamma(H^{++} \rightarrow l^{+}_{\alpha}l^{+}_{\beta})}
	\end{equation}
	
	in the plane of $m_{H^{++}}$ vs $v_{\xi}$ as shown in the Fig.~\ref{Xi_decay_contour}. It can be easily inferred from Fig~\ref{Xi_decay_contour} that if $v_{\xi}$ is a few hundred MeV or more, then $H^{++}$ dominantly decays to $W^{+}W^{+}$. 
	
	\bibliographystyle{JHEP}
	\bibliography{ref}  
\end{document}